\documentclass{emulateapj}
\usepackage{graphicx}

\shorttitle{HV Regions in SN~Ia}
\shortauthors{Marion et al.}

\newcommand{\ca}{\ion{Ca}{2}}
\newcommand{\mg}{\ion{Mg}{2}}
\newcommand{\si}{\ion{Si}{2}}
\newcommand{\sithree}{\ion{Si}{3}}
\newcommand{\s}{\ion{S}{2}}
\newcommand{\fe}{\ion{Fe}{2}}
\newcommand{\oi}{\ion{O}{1}}
\newcommand{\kms}{km~s$^{-1}$}

\newcommand{\wl}{$\lambda$}

\newcommand{\bmax}{$B$-max}

\begin{document} 

\title{High-Velocity Line Forming Regions in the Type I\lowercase{a} Supernova 2009\lowercase{ig}}

\author{G. H. Marion\altaffilmark{1,2}, 
Jozsef Vinko\altaffilmark{2,3}, 
J. Craig Wheeler\altaffilmark{2}, 
Ryan J. Foley\altaffilmark{1,4},\\
Eric Y. Hsiao\altaffilmark{5},
Peter J. Brown\altaffilmark{6}, 
Peter Challis\altaffilmark{1},
Alexei V. Filippenko\altaffilmark{7}, 
Peter Garnavich\altaffilmark{8},\\
Robert P. Kirshner\altaffilmark{1},
Wayne B. Landsman\altaffilmark{9}, 
Jerod T. Parrent\altaffilmark{10,11}, 
Tyler A. Pritchard\altaffilmark{12},\\
Peter W. A. Roming\altaffilmark{12,13},
Jeffrey M. Silverman\altaffilmark{2}, 
and Xiaofeng Wang\altaffilmark{14,6}}

\altaffiltext{1}{Harvard-Smithsonian Center for Astrophysics, 60 Garden Street, Cambridge, MA 02138, USA; \email{gmarion@cfa.harvard.edu}}
\altaffiltext{2}{University of Texas at Austin, 1 University Station C1400, Austin, TX, 78712-0259, USA}
\altaffiltext{3}{Department of Optics and Quantum Electronics, University of Szeged, Dom ter 9, 6720, Szeged, Hungary}
\altaffiltext{4}{Clay Fellow, Harvard-Smithsonian Center for Astrophysics}
\altaffiltext{5}{Carnegie Observatories, Las Campanas Observatory, Colina El Pino, Casilla 601, Chile}
\altaffiltext{6}{George P. and Cynthia Woods Mitchell Institute for Fundamental Physics \& Astronomy, Department of Physics and Astronomy, Texas A\&M University, 4242 AMU, College Station, TX 77843, USA}
\altaffiltext{7}{Department of Astronomy, University of California, Berkeley, CA 94720-3411, USA}
\altaffiltext{8}{Department of Physics, University of Notre Dame, 225 Nieuwland Science Hall, Notre Dame, IN, 46556, USA}
\altaffiltext{9}{Adnet Systems, NASA Goddard Space Flight Center, Greenbelt, MD 20771, USA}
\altaffiltext{10}{Las Cumbres Observatory Global Telescope Network, Goleta, CA 93117, USA}
\altaffiltext{11}{Department of Physics \& Astronomy, Dartmouth College, Hanover, NH 03755, USA}
\altaffiltext{12}{Department of Astronomy \& Astrophysics, Penn State University, 525 Davey Lab, University Park, PA 16802, USA}
\altaffiltext{13}{Space Science \& Engineering Division, Southwest Research Institute, P.O. Drawer 28510, San Antonio, TX 78228-0510, USA}
\altaffiltext{14}{Physics Department and Tsinghua Center for Astrophysics (THCA), Tsinghua University, Beijing 1,00084, China}

%%%%%%%%%%%%%%%%%%%%%%%%%%%%%%%%
\begin{abstract}
We report measurements and analysis of high-velocity ($>$ 20,000 \kms) and photospheric absorption features in a series of spectra of the Type Ia supernova (SN) 2009ig obtained between $-14$~d and $+13$~d with respect to the time of maximum $B$-band luminosity (\bmax).  We identify lines of \si, \sithree, \s, \ca, and \fe\ that produce both high-velocity (HVF) and photospheric-velocity (PVF) absorption features.  SN~2009ig is unusual for the large number of lines with detectable HVF in the spectra, but the light-curve parameters correspond to a slightly overluminous but unexceptional SN~Ia ($M_B = -19.46$ mag and $\Delta m_{15}(B) = 0.90$ mag).  Similarly, the \si\ \wl 6355 velocity at the time of \bmax\ is greater than ``normal" for a SN~Ia, but it is not extreme ($v_{\rm Si}=$13,400 \kms).  The $-14$~d and $-13$~d spectra clearly resolve HVF from \si\ \wl 6355 as separate absorptions from a detached line forming region.  At these very early phases, detached HVF are prevalent in all lines.   From $-12$~d to $-6$~d, HVF and PVF are detected simultaneously, and the two line forming regions maintain a constant separation of about 8,000 \kms.  After $-6$~d all absorption features are PVF.  The observations of SN~2009ig provide a complete picture of the transition from HVF to PVF.  Most SN~Ia show evidence for HVF from multiple lines in spectra obtained before $-10$~d, and we compare the spectra of SN~2009ig to observations of other SN.  We show that each of the unusual line profiles for \si\ \wl 6355 found in early-time spectra of SN~Ia correlate to a specific phase in a common development sequence from HVF to PVF.   
\end{abstract}


%%%%%%%%%%%%%%%%%% Section %%%%%%%%%%%%%%%%%%%%%%
\section{Introduction}

Research on Type Ia supernovae (SN~Ia) has been guided for many years by the general agreement that the progenitors are carbon-oxygen (C/O) white dwarf stars (WD) as first predicted by \citet{hoyle60}.  This hypothesis is based on observational and theoretical evidence that thermonuclear burning of a C/O WD can produce light curves and spectra that are very similar to those observed for SN~Ia.  Beyond that general statement, we lack many of the details concerning the composition of the progenitors and the physics of the explosions.  In order to increase the effectiveness of SN~Ia as distance indicators we must improve our ability to predict the intrinsic brightness of individual events.  Observations that provide new information about the chemical structure deepen our understanding of the explosion physics and may contribute constraints to theoretical models.

Here we identify, measure, and compare high-velocity ($>$ 20,000 \kms) and photospheric-velocity (PVF) absorption features from lines of \si, \sithree, \s, \ca, and \fe\ in a sequence of optical spectra of SN~Ia~2009ig.  The spectra of SN~2009ig at phases earlier than $+3$~d were previously published by \citet{Foley12}.  The phase of each observation is expressed in days with respect to the time of $B$-band maximum light, \bmax\ (Sep 6.0 UT = JD 2,455,080.5; see \S 2.1). 

High-velocity (HVF) absorption features from \ca\ lines are frequently identified in spectra of SN~Ia obtained a week or more before \bmax, but reliable identifications of HVF from other elements are rare.  The presence of HVF from other lines is inferred from unusual PVF line profiles in very early-time spectra.  HVF components of blended \si\ \wl 6355 features are reported by \citet{Mattila05},  \citet{Quimby06}, \citet{Stanishev07}, \citet{Garavini07}, \citet{Tanaka08}, and \citet{Wang09a}.   Evidence for HVF in \fe\ features is discussed by \citet{Branch03}, \citet{Branch05}, \citet{Branch07}, \citet{Hatano99}, and \citet{Mazzali05a}.  

The two earliest spectra of SN~2009ig ($-14.5$~d, $-13.5$~d) clearly resolve high-velocity \si\ \wl 6355 as a separate absorption feature produced in a detached line forming region.  All features identified in the earliest spectra of SN~2009ig are HVF.  The first unambiguous detection of PVF occurs at $-12$~d, and we measure both HVF and PVF velocities from $-12$~d until HVF are no longer detectable at about $-6$~d.   All absorption features in spectra obtained after $-5$~d are PVF.  The early start and dense coverage of our sample allows us to map a complete transition from HVF to PVF in SN~2009ig.

The strongest absorption features in early-time spectra of SN~Ia are the \ca\ H\&K lines and the \ca\ near-infrared triplet (IR3).   Most of the pre-maximum spectra in our SN~2009ig sample include both of these lines, presenting an opportunity to study the development of HVF and PVF in \ca.  \citet{Wang06} and \citet{Patat09} also reported simultaneous observations of HVF and PVF from both \ca\ lines.   

HVF from \ca\ are usually observed as very broad features that extend 15,000--25,000 \kms\ beyond typical photospheric velocities.  They are most often detected in spectra obtained earlier than $-5$~d, but it is not unusual for \ca\ HVF to persist as late as maximum light.  Detections of high-velocity \ca\ in early spectra of SN~Ia are reported and discussed by \citet{Fisher97}, \citet{Hatano99}, \citet{Wang03}, \citet{Gerardy04}, \citet{Thomas04}, \citet{Branch08}, \citet{Tanaka08}, \citet{Branch09}, \citet{Marion09}, and others.  \citet{Childress13b} compare and discuss \ca\ HVF in SN~Ia observed near \bmax.    \citet{Mazzali05a} assert that all spectra of SN~Ia observed more than one week before maximum will exhibit high-velocity \ca.

\citet{Foley12} use the same pre-maximum data of SN~2009ig as we do to identify and discuss differences in line profiles of \ca\ H\&K, \ca\ IR3, and \si\ \wl 6355. They reveal the two-component nature of early \si\ \wl 6355 features and propose that HVF may be ubiquitous in SN~Ia.  Here we focus on the location and composition of HVF line forming regions.  In particular, we examine the characteristics of HVF from lines that are not \ca.  We directly compare the development of \si\ \wl 6355 features in the spectra of SN~2009ig to sequences of early-time spectra of five other SN~Ia.

Velocity measurements show that the HVF and PVF line forming regions remain separated by about 8,000 \kms\ in SN~2009ig. The expansion of SN~Ia is assumed to be homologous, so that measured velocities are proportional to distances from the center of the SN.  Consequently, a line forming region that produces high-velocity absorption features is physically outside of a region that produces lower velocity features.  Homologous expansion ensures that they will remain in the same relative positions.  

The formation of detached HVF requires a region with a localized enhancement of abundance or density. The HVF layer must also be farther from the center of the SN than the photospheric layer.   We are a long way from a consensus for how HVF line forming regions come into existence.  A few of the proposed explanations for HVF are uneven density profiles in the ejecta \citep{Mazzali05a}, a clumpy structure or a thick torus \citep{Tanaka06}, and interaction between the SN ejecta and surrounding material \citep{Gerardy04}.  Evidence for polarization of the HVF in spectra of SN~Ia was first presented by \citet{Wang03} and discussed by \citet{Kasen03}.  If the HVF regions are not spherically symmetric, then viewing angles will influence the observed velocities.  

In \S \ref{obs} we describe the observations of SN~2009ig and the reduction of the data.   A discussion of the methods used for identification and measurement of absorption features is presented in \S \ref{id}.   The characteristics of HVF from specific lines are presented in \S \ref{hv}.  A discussion of PVF and their measurements appears in \S \ref{ps}.  Analysis of the composition and location of HVF line forming regions is found in \S \ref{lfr}. In \S \ref{comp} we compare the results for SN~2009ig to early-time spectra of other SN~Ia. Possible clues to the origins of HVF regions in SN~Ia are discussed in \S \ref{sources}.  Section~\ref{conc} presents a summary and conclusions.

%%%%%%%%%%%%%%%%%% Section %%%%%%%%%%%%%%%%%%%%%%
\section{Observations}
\label{obs}

SN~2009ig in NGC 1015 (redshift $z=0.008770$; NED) was discovered \citep{Kleiser09} by the Lick Observatory Supernova Search \citep[LOSS;][]{Filippenko01} with the 0.76~m Katzman Automatic Imaging Telescope (KAIT) at a magnitude of 17.5 on 2009 Aug. 20.5 (UT dates are used throughout this paper).  The last known nondetection was also by LOSS/KAIT on Aug. 16 to a limiting magnitude of 18.7.   On Aug. 21.1 the first spectrum was obtained with the Asiago 1.82 m telescope by \citet{Navasardyan09}. It revealed characteristics of a SN~Ia very soon after the explosion with a high expansion velocity for \si\ \wl 6355 (24,500 \kms) and no detection of \si\ \wl 5971 or \s\ \wl 5641.  Because SN~2009ig was nearby and the detection was very early, we were presented with a unique opportunity to conduct extensive observations of a young SN~Ia.

\subsection{Photometry}
Photometric data were obtained in the $UBVI$ bands with the 0.8 m Tsinghua-NAOC Telescope (TNT), located at Xinglong Station of the National Astronomical Observatory of China, 180 km from Beijing.   Details of the TNT, detectors, and observing conditions are described by \citet{Wang08}.  All photometric data were reduced using standard IRAF packages.  

\begin{figure}[t]
\center
%\hspace{-0.5cm} % To move left
\includegraphics[width=0.52\textwidth]{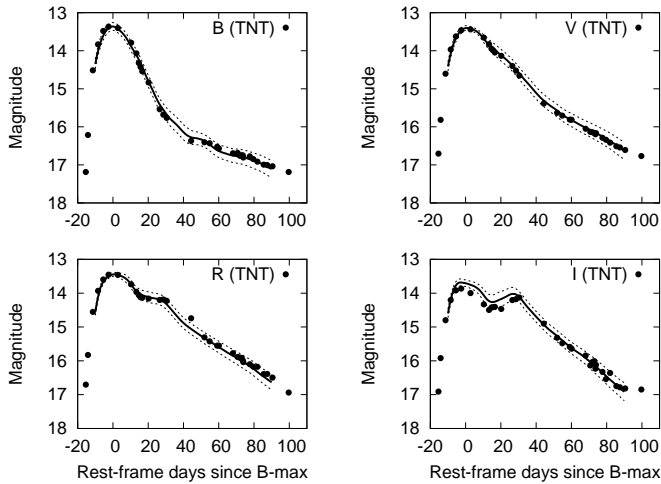}
\caption[]{$BVRI$ light curves of SN~2009ig obtained at the Tsinghua-NAOC Telescope (TNT).  The black solid line is the MLCS2k2 fit and the dotted lines are the $1\sigma$ boundaries.  Template fitting gives $M_B = -19.5$ mag on 2009 Sep. 6.0 with $\Delta m_{15}(B) = 0.90$ mag.  \label{lc}}
\end{figure}

The light curves were fit to MLCS2k2 templates \citep{Jha07}.   The light curves and the best-fit templates are displayed in Figure~\ref{lc}.  We adopted $A_V{\rm (Gal)} = 0.089$ mag for the Milky Way extinction \citep{Schlafly11} and a time of maximum $B$-band brightness of $T(B_{\rm max}) = 55080.50$ MJD (2009 Sep. 6.0).  MLCS2k2 finds $\Delta = -0.24 \pm 0.08$ for the best-fit light-curve parameter, $\mu_0 = 32.82 \pm 0.09$ mag for the distance modulus (assuming H$_0 = 73$ km s$^{-1}$ Mpc$^{-1}$), and $A_V{\rm (host)} = 0.01 \pm 0.01$ mag for the host-galaxy extinction.

From the MLCS2k2 results we find $M_B = -19.46 \pm 0.12$ and $M_V = -19.42 \pm 0.12$ mag on 2009 Sep. 6.0.  The $\Delta = -0.24$ parameter corresponds to a decline-rate value of $\Delta m_{15}(B) = 0.90 \pm 0.07$ mag.  These values suggest that SN~2009ig is slightly more luminous and has a slower decline rate than a ``normal'' SN~Ia, but the $M_B$ and $\Delta m_{15}(B)$ values are not exceptional. 

These parameters are consistent with those of \citet{Foley12}, who used KAIT and \emph{Swift} photometry of SN~2009ig to derive values of $\mu_0 = 32.96 \pm 0.02$ mag and $A_V{\rm (host)} = 0.01 \pm 0.01$ mag.  Small differences in the photometric parameters have no effect on the spectroscopic results presented in this paper.

%%%%****** Figure 1 ******%%%%

%%%%****** Figure 2 ******%%%%
\begin{figure*}[t]
\center
\includegraphics[width=0.76\textwidth]{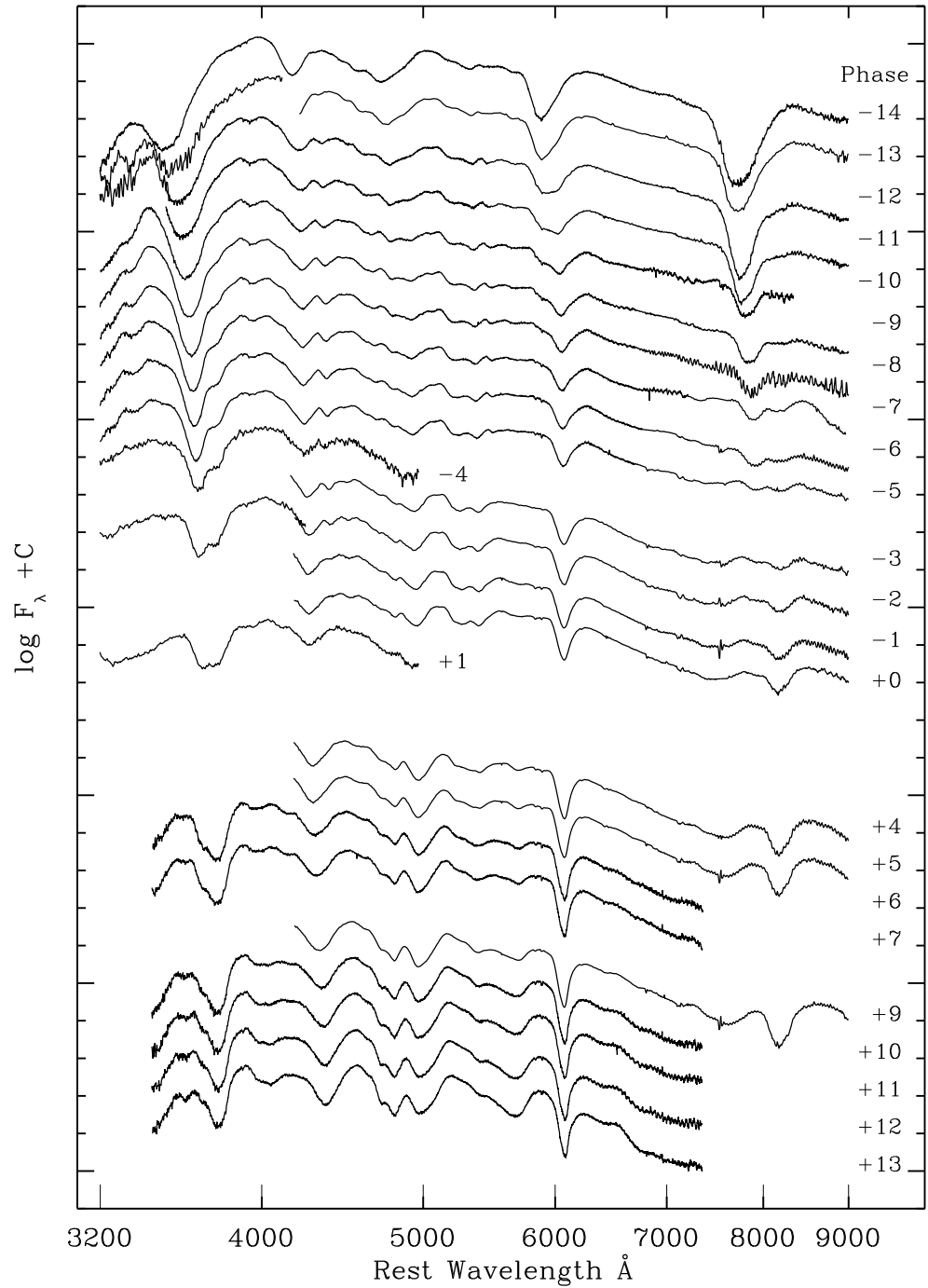}
\caption[]{Spectra of SN~2009ig obtained between $-14$~d (top) and +13~d (bottom) with respect to the time of \bmax\ (2009 Sep. 6.0).   The spectra are positioned in the figure to align the continuum near 4700~\AA\ with the phase as indicated on the ordinate. CCD fringing is apparent at the red end in a few of the spectra.  \ca\ features are strong near 3800~\AA\ for H\&K and near 8,000~\AA\ for IR3, while \si\ \wl 6355 is prominent near 5900~\AA.  HVF components dominate the earliest spectra.  There is a period of transition when both HVF and PVF can be detected from the same lines, and most lines are exclusively PVF after $-6$~d.  Full wavelength coverage is achieved for some of the dates by combining spectra.  See Table~\ref{obstbl} for details of the observations and \S \ref{spec} for an explanation of how the spectra were merged.  \label{fullstack}}
\end{figure*}

%%%%%%%%%%%%%%%%%% Section %%%%%%%%%%%%%%%%%%%%%%
\subsection{Spectroscopy}
\label{spec}

Our complete sample includes 32 optical spectra of SN~2009ig obtained from $-14.5$~d to +12.5~d with respect to the time of \bmax.  Spectra obtained before Sep. 9 (+3 d) were previously published by \citet{Foley12}.  In most cases where the phase is discussed in the text, we round the value to whole days for simplicity.  The \emph{Swift} $U$-grism spectra are included in this list of ``optical" spectra since we only display and analyze the portion of these spectra in the range 3200--4400~\AA.  

Figure~\ref{fullstack} displays spectra obtained on 25 of the 28 nights between $-14$~d and +13 d.  The maximum wavelength range of 3200--9000~\AA\ reaches both the \ca\ H\&K lines (\wl 3945) and the \ca\ IR3 (\wl 8579).  Full wavelength coverage is achieved on every night from $-14$~d to $-5$~d.  

For some of the dates, observations of SN~2009ig that cover slightly different wavelengths occurred at two different telescopes within a few hours of each other.  We combine these data to create a single continuous spectrum by trimming the data from one source at a specific wavelength and continuing from that wavelength with data from the other source without overlap.  The transition points are selected to optimize the signal-to-noise ratio for the combined spectrum.  Spectral features in SN~Ia change rapidly at early times, but a daily cadence is sufficient to trace the development of individual features.  The use of a single spectrum per day makes it easier to measure and display the data.  A complete list of the observational details can be found in Table~\ref{obstbl}, including the wavelengths at which the spectra were joined.  

For the ground-based spectroscopy, the slit was generally oriented along the parallactic angle to minimize differential slit losses caused by atmospheric dispersion \citep{Filippenko82}.   SN~2009ig was observed using the Shane 3 m telescope at Lick Observatory with the Kast spectrograph \citep{Miller93} on Aug. 22, 24, 25, 27, and 28.  An additional spectrum was obtained on Aug. 22 at the 10 m Keck I telescope with the Low Resolution Imaging Spectrometer \citep[LRIS;][]{Oke95} equipped with an atmospheric dispersion corrector.  The Lick/Kast spectra used a 600/4310 grism on the blue side and a 300/7500 grating on the red side with a $2''$ wide slit. The Keck/LRIS spectrum was obtained with a 400/3400 grism on the blue side and a 600/7500 grating on the red side using a $1''$ wide slit.  

The 9.2 m Hobby-Eberly Telescope \citep[HET;][]{Ramsey98} with the Marcario Low-Resolution Spectrograph \citep[LRS;][]{Hill98} was used to observe SN~2009ig on Aug. 23 ($-13$~d).  Additional HET observations were made on Aug. 29, 30, and 31, as well as on Sep. 2, 3, 4, 5, 9, 10, and 14.  The HET/LRS spectra have an effective wavelength range of 4400--9200~\AA.  Reduction of the HET/LRS data was also conducted with standard IRAF procedures.  

Low-dispersion spectra of SN~2009ig were obtained on Aug. 26, 27, 28, 29, and 30 using the 6.5 m MMT with the Blue Channel spectrograph \citep{Schmidt89}, covering the wavelength range 3200--8200~\AA.  CCD processing and spectrum extraction for these observations were completed with IRAF and the data were extracted with the optimal algorithm of \citet{Horne86}.   After the wavelength calibration was derived from low-order polynomial fits to calibration-lamp spectra, additional small adjustments were applied by cross-correlating a template sky spectrum to the night-sky lines that were extracted along with the SN. IDL routines were used to flux calibrate the data and to remove telluric lines \citep{Wade88, Matheson00, Foley03}.

The \emph{Swift} satellite observed SN~2009ig with the UV grism mode of the Ultraviolet/Optical Telescope \citep[UVOT;][]{Roming05} onboard the spacecraft \citep{Gehrels04}.  \emph{Swift} data were obtained on Aug. 23, 25, and 27, and on Sep. 1, 3, and 6.  We do not use all of the \emph{Swift} data for our analysis, but the complete sequence of \emph{Swift} spectra and photometric measurements was published by \citet{Foley12}.  \emph{Swift} data were extracted and reduced using the ``uvotimgrism'' package of FTOOLS.  The effective wavelength coverage is 1700--4900~\AA, but in this analysis we use only a small portion of the \emph{Swift} spectra covering the wavelength region 3200--4400~\AA\ that contains the \ca\ H\&K feature.  

%%%%****** Figure 3 ******%%%%
\begin{figure}[t]
\center
\includegraphics[width=0.5\textwidth]{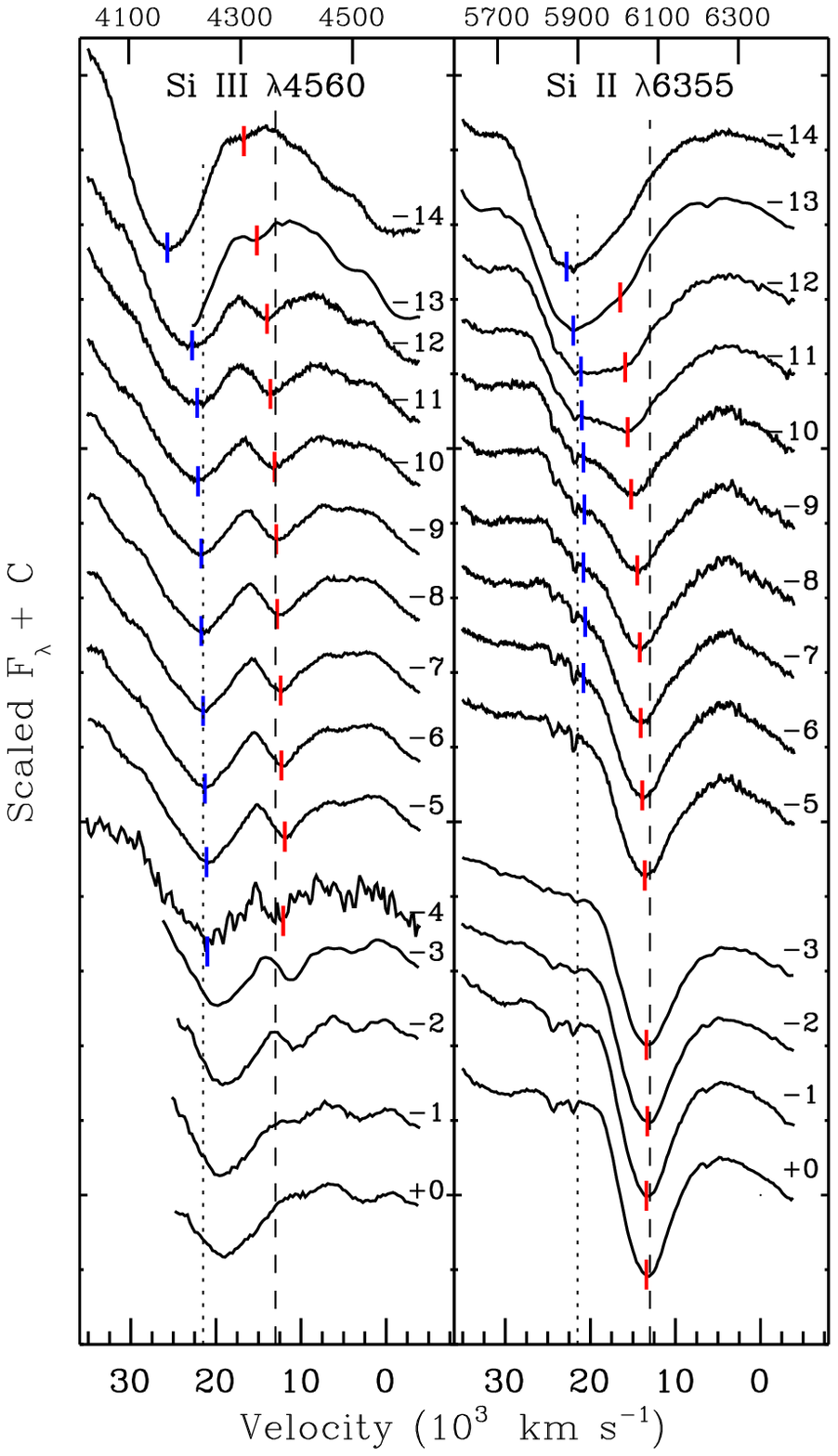}
\caption[]{HVF and PVF for \sithree\ \wl 4560 (left) and \si\ \wl~6355 (right) in SN 2009ig are displayed in velocity space (blueshifted velocities are expressed with positive values). Rest wavelengths are indicated on the top abscissa. The spectra were obtained between $-14$~d and \bmax, and the phase of each each spectrum is marked near the red end.  The line profiles for \si\ \wl 6355 in the earliest spectra clearly require separate HVF and PVF components.  The phases from $-12$~d to $-8$~d reveal contributions from both HVF and PVF.  Narrow \ion{Na}{1}~D features are visible in the \si\ \wl~6355 profiles.  HVF \sithree\ \wl 4560 is blended with PVF \mg\ \wl 4481 beginning about $-10$~d.   The velocities measured for each feature are indicated by a blue hashmark for HVF and a red hashmark for PVF.  The dotted line at 21,500~\kms\ and the dashed line at 13,000~\kms\ are the same in all figures.  \label{pvsi}}
\end{figure}

%%%%****** Figure 4 ******%%%%
\begin{figure}[t]
\center
\includegraphics[width=0.5\textwidth]{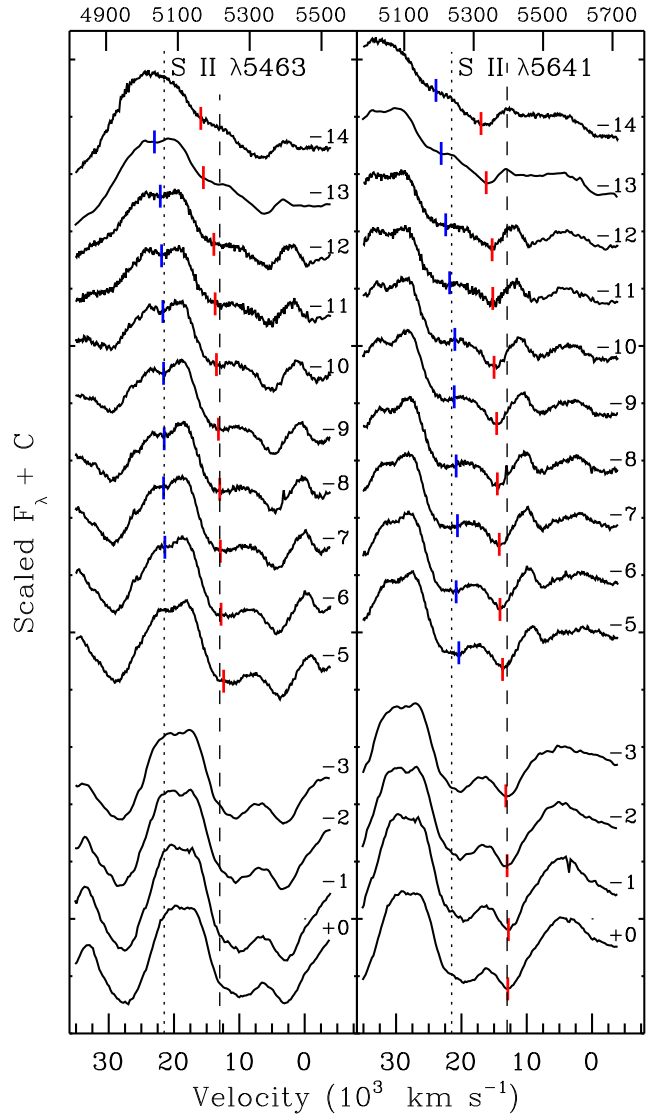}
\caption[]{HVF and PVF for \s\ \wl 5453 (left) and  \s\ \wl 5641 (right) are marked on a sequence of spectra of SN~2009ig obtained between $-14$~d and +0 d. Rest wavelengths are indicated on the top abscissa. The phase of each spectrum is marked near the red end.  HVF are strongest before $-10$~d and PVF are stronger after $-8$~d.  The measured velocities are indicated by a blue mark for HVF and a red mark for PVF.   The PVF in \s\ \wl 5453 are compromised by blending after $-5$ d and we do not measure the velocities (see \S \ref{ps}). The dotted line at 21,500~\kms\ and the dashed line at 13,000~\kms\ are the same in all figures.  \label{pvs}}
\end{figure}

%%%%****** Figure 5 ******%%%%
\begin{figure}[t]
\center
\includegraphics[width=0.5\textwidth]{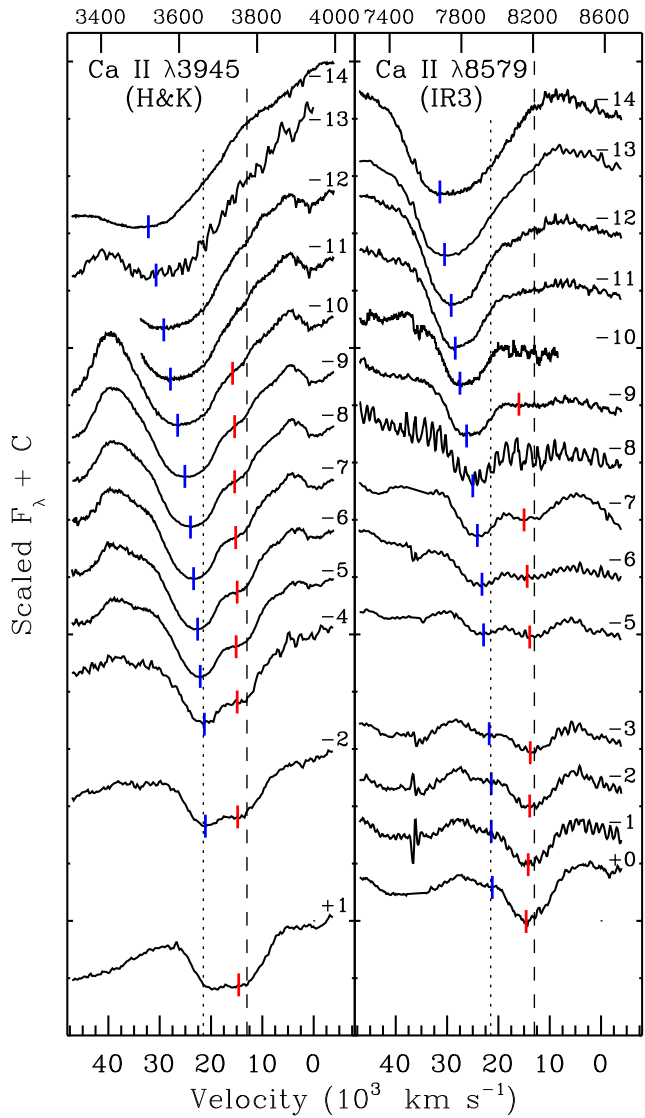}
\caption[]{HVF and PVF for \ca\ H\&K \wl 3945 (left) and \ca\ IR3 \wl 8579 (right) in SN 2009ig are displayed in velocity space (blueshifted velocities are expressed with positive values). Rest wavelengths are indicated on the top abscissa. The spectra were obtained between $-14$~d and $+1$~d and the phase of each each spectrum is marked near the red end.  HVF velocities for \ca\ are higher than for other lines in the early spectra (compare offsets from the dotted line), but they eventually decline to the mean HVF velocity for all lines (dotted line).    \ca\ HVF are detected for a few days longer than HVF for other lines, and the PVF appear later.  Comparing the features by phase suggests that the HVF component of \ca\ H\&K \wl 3945 is exaggerated by a strengthening contribution from PVF \si\ \wl 3858 beginning about $-10$~d.  The measured velocities are indicated by a blue mark for HVF and a red mark for PVF.  The dotted line at 21,500~\kms\ and the dashed line at 13,000~\kms\ are the same in all figures. \label{pvca}}
\end{figure}

%%%%****** Figure 6 ******%%%%
\begin{figure}[t]
\center
\includegraphics[width=0.5\textwidth]{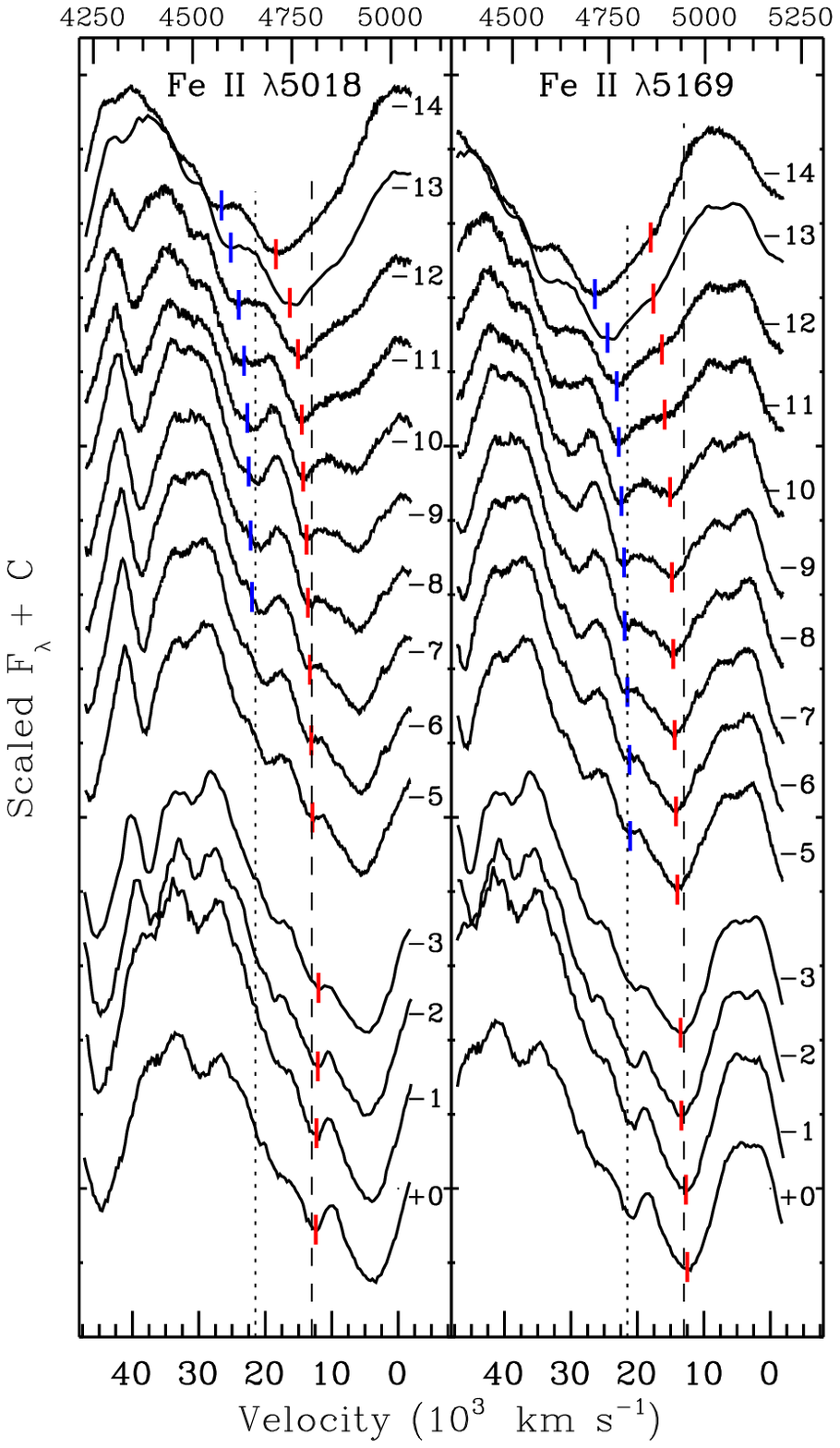}
\caption[]{HVF and PVF for \fe\ \wl 5018 (left) and \fe\ \wl 5169 (right) are marked on a sequence of spectra of SN~2009ig obtained between $-14$~d and +0 d.  Rest wavelengths are indicated on the top abscissa. The phase of each spectrum is marked at the red end.  The dashed line in the velocity space of  \fe\ \wl 5018 and the dotted line in velocity space for HVF \fe\ \wl 5169 are in the same physical location.  The feature at that position is produced by HVF only in the first two spectra, by a blend of PVF and HVF from about $12$~d to $-6$~d, and by PVF only after $-6$~d.  Comparing offsets from the dotted lines shows that HVF of \fe\ are deeper, wider, and at higher velocities than HVF of \si\ (Fig.~\ref{pvsi}) and \s\ (Fig.\ref{pvs}).  The measured velocities are indicated by a blue mark for HVF and a red mark for PVF.  The dotted line at 21,500~\kms\ and the dashed line at 13,000~\kms\ are the same in all figures.  \label{pvfe}}
\end{figure}

%%%%%%%%%%%%%%%%%% Section %%%%%%%%%%%%%%%%%%%%%%
\section{Line Identifications and Velocity Measurements}
\label{id}

\placefigure{pvsi}
\placefigure{pvs}
\placefigure{pvca}
\placefigure{pvfe}

Eight lines have been identified that produce both HVF and PVF absorptions in spectra of SN~2009ig.  The features associated with these lines are displayed by phase and velocity space in four figures: \sithree\ \wl 4560 and \si\ \wl~6355 (Figure~\ref{pvsi}),  \s\ \wl\wl\ 5453, 5641 (Figure~\ref{pvs}),  \ca\ H\&K \wl 3945 and \ca\ IR3 \wl 8579 (Figure~\ref{pvca}), and \fe\ \wl\wl\ 5018, 5169 (Figure~\ref{pvfe}).   The velocity measurements at each phase are indicated by a blue hashmark for HVF and a red hashmark for PVF.  To facilitate comparison of the components between features, all of the figures also have a dotted line plotted at 21,500~\kms\ and a dashed line at 13,000~\kms.  These velocities were selected to be representative of the HVF at $-10$~d and the PVF near \bmax.   A detailed discussion of individual features is found in \S \ref{hv} (HVF) and \S \ref{ps} (PVF).

The HVF and PVF of \si\ \wl 6355 are easily identified and relatively free from blending.  These characteristics make them useful templates that guide efforts to identify absorption features associated with other lines.  We compare the velocities, the phases of initial and final detections for HVF and PVF, the relative strengths of the HVF and PVF components at a given phase, and the velocity separation between the HVF and PVF.  The presence of simultaneous HVF and PVF from several lines increases the number of blended features in the early-time spectra of SN~2009ig.  Velocities and line profiles of individual features are measured directly from the flux-calibrated spectra rather than a flattened continuum because the shapes and slopes of the local continua vary widely in different regions of each spectrum.

In the velocity figures, the flux is scaled to make it easier to see details of the absorption features.  A representative flux ratio is maintained between the two panels in each figure in order to compare the relative strengths of features from similar lines, but it is not possible to maintain a consistent scale between the figures.  The velocity range for the \ca\ and \fe\ plots 48,000 to $-$8,000 \kms. For \si\ and \s\ velocity plots the range is 36,000 to $-$8,000 \kms.   Throughout this paper we reference blueshifted velocities using positive values.  

%%%%****** Figure 7 ******%%%%
\begin{figure}[t]
\center
%\hspace{-0.5cm} % To move left
\includegraphics[width=0.52\textwidth]{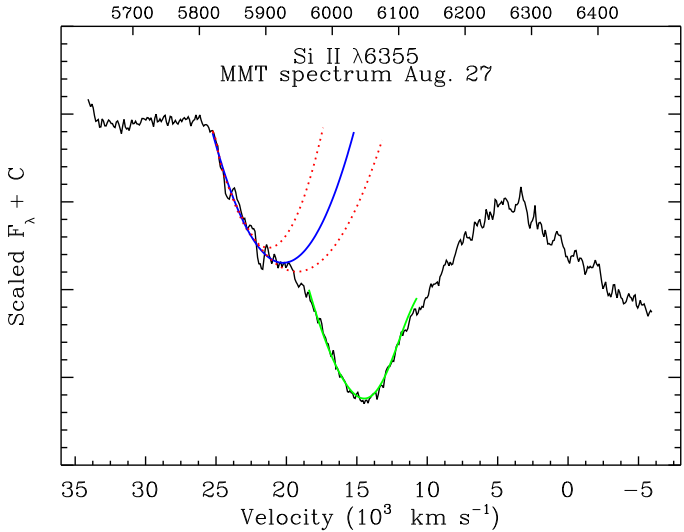}
\caption[]{Gaussian profiles are used to measure blended absorption features if one of the components does not produce a distinct minimum.  This example shows the \si\ \wl 6355 feature in the MMT spectrum of SN~2009ig obtained on Aug. 27 ($-9$~d). Rest wavelengths are indicated on the top abscissa. The PVF component is dominant at this phase, but there is clear evidence for a contribution from the HVF component.  The solid blue line shows that a Gaussian profile with a minimum at 20,700 \kms\ is a good fit to the data.  The red dotted lines form Gaussians with minima that have been shifted to 19,700 and 21,700 \kms.  The red lines do not fit the data as effectively as the blue line.  The photospheric component is fit with a Gaussian (solid green line) having a minimum at 14,400 \kms. \label{gfit}}
\end{figure}

At many phases one of the components may lack a distinct minimum but still makes an obvious contribution to the overall line profile.   For example, the right panel of Figure~\ref{pvsi} shows that the PVF component from \si\ \wl 6355 at $-12$~d produces only a small indentation on the red side of the larger absorption feature dominated by HV.  As time passes the strength of the HVF signal diminishes and the PVF component becomes stronger.  At $-9$~d the PVF absorption is dominant and the HVF component forms a small but distinct indentation on the blue side of the PVF absorption feature.  We use this spectrum to illustrate a method for measuring velocities from absorption components that form incomplete features.

Figure \ref{gfit} shows the \si\ \wl 6355 feature from the August 27 spectrum ($-9$~d).  The photospheric component is simply fit by a Gaussian (solid green line) having a minimum at 14,400 \kms.   The data in the region of the indentation created by the HVF component are fit with a Gaussian profile having a minimum at 20,700 \kms\ (solid blue line).  The dotted red lines show alternative Gaussians with minima that have been shifted by $\pm$ 1,000 \kms\ from the minimum of the blue line.  Neither of the line profiles plotted in red fit the data as well as the blue line.  The optimal position for the Gaussian is determined by changing the minimum in steps of 100 \kms\ and inspecting the results by eye.  It is nearly always easy to determine when the Gaussian profile deviates from the actual data by more than a few hundred \kms.  Uncertainties due to this measurement technique are 100--1,000 \kms\ depending on the quality of the data and how well the minima are defined.   The measurements in Tables~\ref{hvtbl} (HVF) and \ref{pstbl} (PVF) were obtained by this method.    Figure~\ref{pvsi} illustrates that for \si\ \wl 6355, uncertainties will be higher for HVF measured after $-9$~d and for PVF measured before $-11$~d.

The SYNOW code \citep{Branch03} is another tool for line identification.  SYNOW creates model spectra that can compare relative line strengths and positions by wavelength.  SYNOW makes it possible to explore the contributions of individual ions to a complete spectrum, and it is particularly helpful for distinguishing between the relative contributions of different lines in a blended feature.

%%%%%%%%%%%%%%%%%% Section %%%%%%%%%%%%%%%%%%%%%%
\section{High-Velocity Absorption Features\\
 ($>$ 20,000 \kms)}
\label{hv}

In most SN~Ia, \ca\ produces the only detectable HVF.  In SN~2009ig, eight lines from five ions produce distinct HVF in the early spectra: \sithree\ \wl 4560 and \si\ \wl 6355 (Figure~\ref{pvsi}), \s\ \wl\wl 5453, 5641 (Figure~\ref{pvs}), and \fe\ \wl\wl 5018, 5169 (Figure~\ref{pvfe}).  There are also strong suggestions of HVF from several other lines that are associated with blended absorption features for which we are unable to confirm the contributions of individual lines.

HVF are clearly identified from all lines in the $-14$~d spectrum of SN~2009ig, and HVF remain strong through $-12$~d.   The strength of the HVF absorptions weaken with time, and in most lines HVF are no longer detectable after $-6$~d.  Beginning at about $-10$~d, many of the HVF are affected by blends with PVF that are increasing in strength.  

The highest velocities for features from each line are measured in the earliest spectrum at $-14$~d.  For \ca, $v_{\rm max} \approx$ 32,600 \kms, followed by \fe\ at 26,600 \kms, \sithree\ at 25,700 \kms, \s\ at 23,900 \kms, and \si\ at 22,800 \kms.   This range of velocities for the HVF is apparently caused by differences in the effective opacities of the lines at that phase.  By $-12$~d,  the HVF velocities for all lines except \ca\ are grouped closely together with a mean of 22,600 \kms.   HVF velocities for all lines converge near 21,500 \kms\ by $-6$~d.

\subsection{HVF from \si, \sithree, and \s}

The HVF of \si\ \wl 6355 provides a useful guide for the behavior of HVF in general because it is essentially unblended.  The HVF component dominates the \si\ profile from $-14$~d to $-12$~d at velocities that are about 1,000 \kms\ lower than \s\ and about 2,000 \kms\ lower than \fe.  Both HVF and PVF contribute to the \si\ \wl 6355 line profile between $-12$~d and $-7$~d, and HVF are no longer detected from \si\ after $-6$~d.   Narrow \ion{Na}{1}~D features are visible in the spectra at about 23,500 \kms, which is zero velocity for Na~D, and at 24,200 \kms, which is the velocity of the host galaxy, NGC~1015.

\si\ has many other lines that often produce detectable features in spectra of SN~Ia.  Possible evidence exists for HVF components from \si\ \wl\wl 3858, 4130, 5051, and 5972, but the features are compromised by blending.

The data do not include sufficiently short wavelengths to cover the HVF of \sithree\ \wl 4560 at $-13$~d, but this feature is strong at $-14$~d and $-12$~d.  Contributions from PVF \mg\ \wl 4481 and PVF \ion{Fe}{3} \wl 4407 begin to affect HVF \sithree\ \wl 4560 by $-11$~d.  We measure HVF \sithree\ until $-4$~d, when the absorption minimum shifts abruptly to the red, indicating that the HVF contribution has ended.  SYNOW models suggest that the PVF \ion{Fe}{3} component is stronger than PVF \mg\ from $-10$~d until about $-3$~d.   

Absorption features in the spectra can possibly be associated with several other doubly ionized lines.  Possible HVF are found for \sithree\ \wl 5740, \ion{S}{3} \wl 4264, and \ion{Fe}{3} \wl\wl 4407, 5129, but as for many other tentative associations, blending prevents definitive identifications.

Figure~\ref{pvs} shows that HVF for both \s\ \wl 5453 and \s\ \wl 5641 are stronger at $-12$~d than they are at $-14$~d.  HVF velocities for \s\ are between the velocities of HVF \fe\ and HVF \si.  The figure also shows that the \s\ \wl 5641 HVF and the \s\ \wl 5453 PVF form a blended feature.  We identify the red component of this blend as HVF \s\ \wl 5641.  The HVF is strongest from $-14$~d to $-12$~d, contributions from both HVF and PVF are evident from $-10$~d to $-5$~d, and weak absorptions remain near the locations of both \s\ HVF as late as $-3$~d.  We stop measuring HVF \s\ \wl 5453 after $-6$~d and HVF \s\ \wl 5641 after $-5$~d.  There is possible evidence in the spectra for HVF from \s\ \wl 5032 and \wl 5429.

HVF are not detected from carbon, oxygen, or magnesium, so there is no evidence for unburned fuel or carbon burning products in the HVF region of SN~2009ig.  If the \mg\ \wl 4481 line produced a HVF at $-14$~d, it would have a velocity of about 24,000 \kms, and it  would appear at about 29,000 \kms\ in the velocity space of \sithree\ \wl 4560 (left panel of Figure~\ref{pvsi}).   At $-10$~d, the \mg\ HVF velocity would be about 21,500 \kms, corresponding to 26,500 \kms\ in the figure.  The spectra are steeply sloped toward the red in this region, which makes it more difficult to discern small absorptions, but careful inspection reveals no evidence of absorption features that can be associated with HVF from \mg.  

\subsection{HVF from \ca\ and \fe}
\label{ca}

The velocity plots and tables show that HVF of \ca\ and \fe\ are wider and deeper, with absorption minima at higher velocities than the HVF of other lines.   \citet{Foley12} observed that HVF of both \ca\ lines are stronger and have higher velocities than HVF of any other lines in the spectra of SN~2009ig.   

The \fe\ \wl 5169 HVF produces a very strong absorption from $-14$~d to $-11$~d that is nearly as broad and deep as the \ca\ HVF.  The feature is subsequently affected by PVF \fe\ \wl 5018, PVF \si\ \wl 5051, and PVF \s\ \wl 5032 beginning about $-10$~d.  We stop measuring HVF \fe\ \wl 5169 after the PVF blend becomes dominant at $-5$~d.  

HVF \fe\ \wl 5018 is the strongest part of a blend with the HVF components of \s\ \wl 5032 and \si\ \wl 5051 from $-14$~d to about $-10$~d.  This identification is confirmed by the close agreement of the measured velocities of HVF \fe\ \wl 5169 and HVF \fe\ \wl 5018.  This feature loses a distinct minimum after $-7$~d and we stop measuring HVF \fe\ \wl 5018 at that point.  Possible evidence exists for HVF components from \fe\ \wl\wl 4924, 5129. 

Figure~\ref{pvca} displays the \ca\ H\&K blend \wl 3935 (left) and the \ca\ IR3 \wl 8579 (right).  At $-14$~d, the absorption minima for both \ca\ blends are near 32,000 \kms\ and the features have blue wings that extend beyond 40,000 \kms.   Simultaneous observations of \ca\ H\&K and IR3 are of particular interest as they provide a rare opportunity to compare the evolution of these features.  As shown in Figure \ref{pvca} and Table \ref{hvtbl}, the measured velocities and rates of change are very similar for HVF from both of the strong \ca\ blends.  Although \ca\ HVF velocities are about 7,000 \kms\ higher than \fe\ HVF and nearly 10,000 \kms\ higher than \si\ HVF at $-14$~d, they have a significantly faster decline rate, and the HVF velocities for all lines converge near 21,500 \kms\ at $-6$~d. 

HVF from the \ca\ IR3 are unblended, and they are weak but detectable until \bmax.  That is about 5 days after the last HVF detection from other lines.  The HVF component of \ca\ H\&K \wl 3945 is blended with PVF \si\ \wl 3858.  Side by side comparison of the \ca\ H\&K and IR3 line profiles suggests that the HVF from \ca\ H\&K are broadened by \si\ PVF beginning about $-10$~d and the PVF influence gets stronger with time. We find that $-2$~d is the final phase at which the absorption minimum of the Ca/Si blend is consistent with the HVF of \ca\ IR3, so we stop measuring \ca\ H\&K HVF after that phase.

\citet{Foley13} demonstrated that there are conditions under which the absorption feature attributed to HVF \ca\ H\&K can be produced by a combination of \si\ and a particular density profile for \ca.  \citet{Wang06} used polarization data in a $-6$~d spectrum from SN~2004dt to suggest that \si\ provides the dominant contribution to this blend at $-6$~d.

\citet{Childress13b} compare the average strength of the \ca\ HVF component in spectra from SN~Ia at \bmax.  They find that slowly declining SN~Ia produce either high photospheric velocities or strong HVF \ca\ close to \bmax, but not both.  SN~2009ig fits this pattern with $v_{\rm Si} =$ 13,400 \kms, while the \ca\ IR3 series in Figure~\ref{pvca} shows that HVF \ca\ is very weak at \bmax.

%%%%%%%%%%%%%%%%%% Section %%%%%%%%%%%%%%%%%%%%%%
\section{Photospheric-Velocity Absorption Features}
\label{ps}

We use velocity offsets from PVF and their relative strengths by phase to confirm the identifications of HVF in spectra of SN~2009ig.  PVF are identified for each of the same lines that produce HVF detections: \sithree\ \wl 4560 and \si\ \wl 6355 (Figure~\ref{pvsi}), \s\ \wl\wl 5453, 5641 (Figure~\ref{pvs}), and \fe\ \wl\wl 5018, 5169 (Figure~\ref{pvfe}).  PVF velocity measurements are in Table~\ref{pstbl}.  

There are weak suggestions of PVF for some lines in the $-14$~d spectrum of SN~2009ig, but the first unambiguous detections of PVF are at $-12$~d.   \citet{Foley12} note that $-12$~d is the phase at which the rate of change slows down for $B-V$ colors.  PVF become progressively stronger with time, and the PVF of most lines are stronger than the HVF by about $-10$~d.   At $-6$~d the PVF components dominate all blended absorption features with little or no influence from the HVF.   

PVF velocities in SN~2009ig decline steeply at first and then more gradually.  They become essentially constant after $-4$~d.  This behavior is consistent with the typical pattern for PVF velocities in SN~Ia \citep{Foley11,Silverman13}.  The PVF velocities of SN~2009ig are about 2,000 \kms\ greater than velocities in a ``normal" SN~Ia, but they are not exceptional. We measure $v_{\rm Si}=$ 13,400 \kms, which places SN~2009ig in the \emph{HV} group defined by \citet{Wang09b}.   \citet{Foley12} show that $v_{\rm Si} <$ 13,500 \kms\ for $\sim 85$\% of SN~Ia having $\Delta m_{15}(B) \le 1.5$ mag.  The change in \si\ \wl 6355 velocity between +0~d and +10~d is $\sim40$ \kms\ per day.   That puts SN~2009ig in the \emph{Low Velocity Gradient} group defined by \citet{Benetti05}.

\subsection{Individual PVF}

PVF \si\ \wl 3858 is blended with HVF \ca\ H\&K.  The \si\ contribution is significant as early as $-10$~d, but it is not clear when the \si\ PVF become strong enough to define the minimum in this feature.  We do not take velocity measurements from this line.  PVF \si\ \wl 4130 is the probable source of the absorption visible as a small notch near 4,000 \AA\ in the P Cygni emission peak from \ca\ H\&K.  HVF \ion{S}{3} \wl 4264 is a likely contributor to this feature.  PVF of \si\ \wl\wl 5051, 5972 are possibly identified.

Most of the \fe\ PVF are strongly affected by blending with other PVF.  PVF \fe\ \wl 5018 (left panel Figure~\ref{pvfe}) is blended with PVF \si\ \wl 5051 and PVF \s\ \wl 5032.   The \si\ and \s\ lines may pull the observed minimum of this feature to the red since the PVF velocities for \fe\ \wl 5018  before \bmax\ are 1,000--1,500 \kms\ lower than PVF \fe\ \wl 5169. 

The strong absorption feature near 4950~\AA\ is formed by PVF \fe\ \wl 5169 with contributions from PVF \fe\ \wl 5266 and PVF \ion{Fe}{3} \wl 5129.   The blend of PVF \fe\ \wl 5018, \si\ \wl 5051, and \s\ \wl 5032 is observed near 4800~\AA, and this feature continues to grow stronger after \bmax.  The P Cygni emission component of this feature pushes the minimum for PVF \fe\ \wl 5169 to velocities that are about 1,000 \kms\ lower than PVF \fe\ \wl 5018 and other PVF lines after \bmax. 

PVF \s\ \wl 5453 is blended with HVF \s\ \wl 5641 from $-12$~d to $-9$~d.  We stop measuring PVF \s\ \wl 5453 after $-5$~d when the velocity abruptly drops from 12,400 \kms\ at $-5$~d to 10,400 \kms\ at $-3$~d.  That is not consistent with the other PVF lines at this phase including the other \s\ line at \wl 5641.  The displacement of the PVF \wl 5453 feature suggests that it is also moved to the red by P Cygni emission from the strong blend of Fe-group lines observed at 4950~\AA.

PVF from lines of doubly ionized ions are weak at $-14$~d and $-13$~d, become significantly stronger at $-12$~d, and disappear at about $-3$~d.  Figure~\ref{pvsi} shows this pattern for PVF \sithree\ \wl 4560.  Other possible but unconfirmed identifications of PVF from twice-ionized lines show similar behavior: \sithree\ \wl 5740, \ion{S}{3} \wl 4264, and \ion{Fe}{3} \wl\wl 4407, 5129.   The HVF components of \s\ lines also appear to be stronger at $-12$~d than at $-14$~d.

PVF \mg\ \wl 4481 is a major contributor to the very strong absorption found near 4220~\AA\ in a blend with PVF \ion{Fe}{3} \wl 4407 and HVF \sithree\ \wl 4560.  The left panel of Figure~\ref{pvsi} shows the minimum of this feature near the dotted line at 21,500 \kms\ until $-3$~d.  A \mg\ PVF with a velocity of 14,000 \kms\ would appear at about 19,000 \kms\ on the velocity scale of \sithree\ \wl 4560.  SYNOW models show a better fit to the data when the PVF \ion{Fe}{3} component is stronger than \mg\ from $-10$~d until about $-3$~d.  The SYNOW result is consistent with the abrupt shift of this feature to the red at $-3$~d that aligns the minimum with characteristic PVF velocities for \mg\ \wl 4481.  

The evidence for PVF from unburned C and O in SN~2009ig is inconclusive.  \ion{C}{2} \wl 4743 makes a reasonable fit to absorption features for both HVF near 4370~\AA\ and PVF near 4480~\AA\ (Figure~\ref{pvsi}, wavelength axis on top).  If, however, \ion{C}{2} \wl 4743 is the source of these features, then \ion{C}{2} \wl 6580 should also be easily detected.  The PVF component of \ion{C}{2} \wl 6580 would appear near 3000 \kms\ in the velocity space of \si\ \wl 6355. \citet{Foley12} cite work with SYNOW models \citep{Parrent11} that finds \ion{C}{2} \wl 6580 to be present in the early-time spectra of SN~2009ig.  \ion{C}{2} is purported to cause a flattening of the red wing in the line profile of \si\ \wl 6355 from $-14$~d to $-12$~d.  It may be that including \ion{C}{2} improves the SYNOW fit to the line profile, but we do not find any absorption features at this location; the tiny notch at 6200 \AA\ is telluric.  

\oi\ \wl 7773 develops a broad PVF line beginning about $-3$~d, but it is much weaker than PVF \oi\ found in many SN~Ia.  The features near 7400 \AA\ in the spectra from $-3$~d to +9~d correspond to PVF of \oi\ \wl 7773 at 11,000--12,000 \kms, consistent with other PVF velocities those phases.

%%%%****** Figure 8 ******%%%%
\begin{figure}[t]
\center
%\hspace{-0.2cm} % To move left
\includegraphics[width=0.45\textwidth]{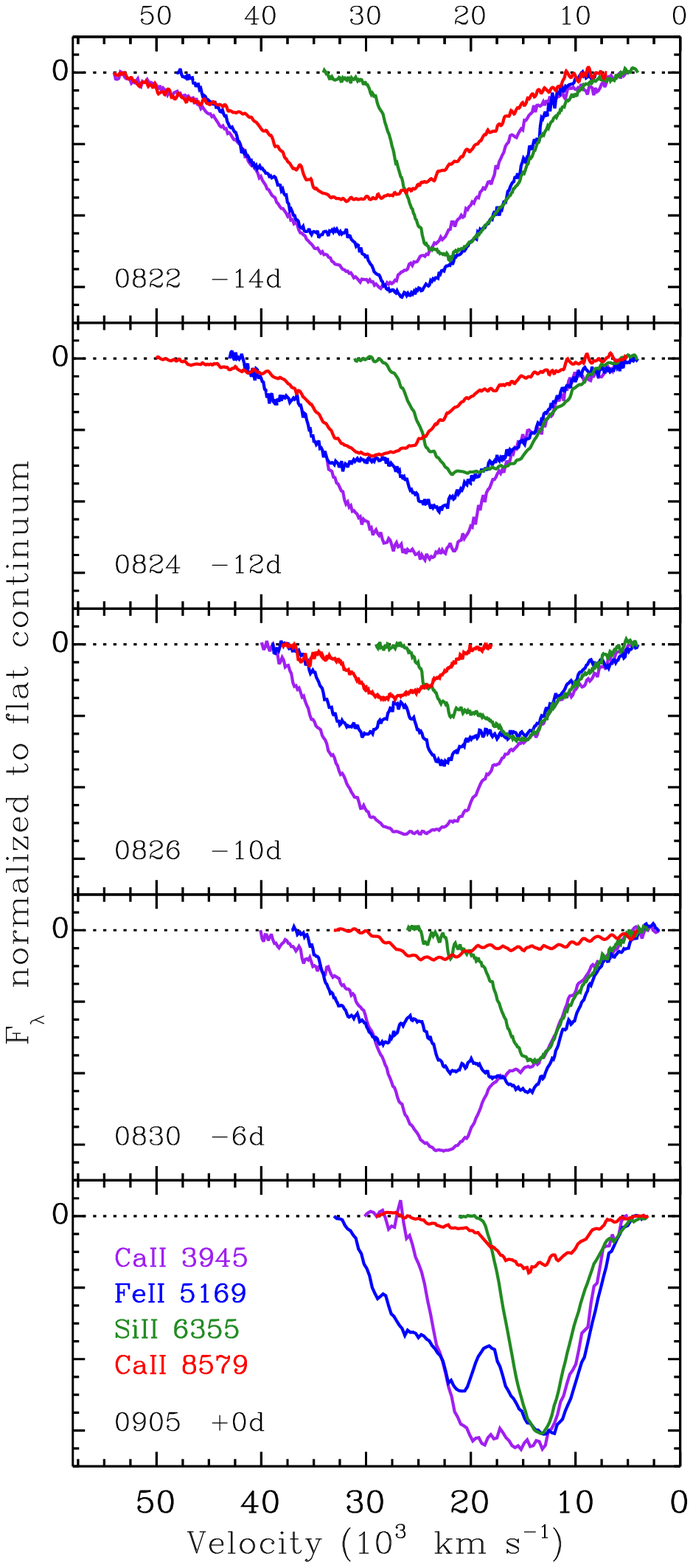}
\caption[]{HVF of \ca\ H\&K (purple), \ca\ IR3 (red), \fe\ \wl 5169 (blue), and \si\ \wl 6355 (green) are plotted together in velocity space at phases $-14$~d, $-12$~d, $-10$~d, $-6$~d, and $+0$~d (top to bottom).  Both HVF of \ca\ occupy the same physical space at $-14$~d.  The \fe\ HVF has a similar profile with an offset of about 7,000 \kms.  The HVF of \fe\ \wl 5018 forms the notch near 35,000 \kms.  At $-14$~d, the \si\ HVF is confined to the lower velocity half of the \fe\ region.  At $-12$~d, HVF remain strong from all lines and weak PVF are detected from \si\ and \fe.   Weakening HVF of \si\ and \fe\ are still detected at $-10$~d, and \ca\ HVF remain strong.  At this phase HVF \fe\ \wl 5169 is affected by blending with PVF \fe\ \wl 5018, and HVF \ca\ H\&K is blended with \si\ \wl 3858.   At $-6$~d, the only evidence for HVF is very weak \ca, and at $+0$~d, all absorptions are PVF.  The HVF of \si\ ($-14$~d) are asymmetric with steeper blue sides while PVF of SiII ($-6$~d,$+0$~d) are nearly symmetric.}{ \label{area4}}
\end{figure}

%%%%****** Figure 9 ******%%%%
\begin{figure*}[t]
\center
\includegraphics[width=1.0\textwidth]{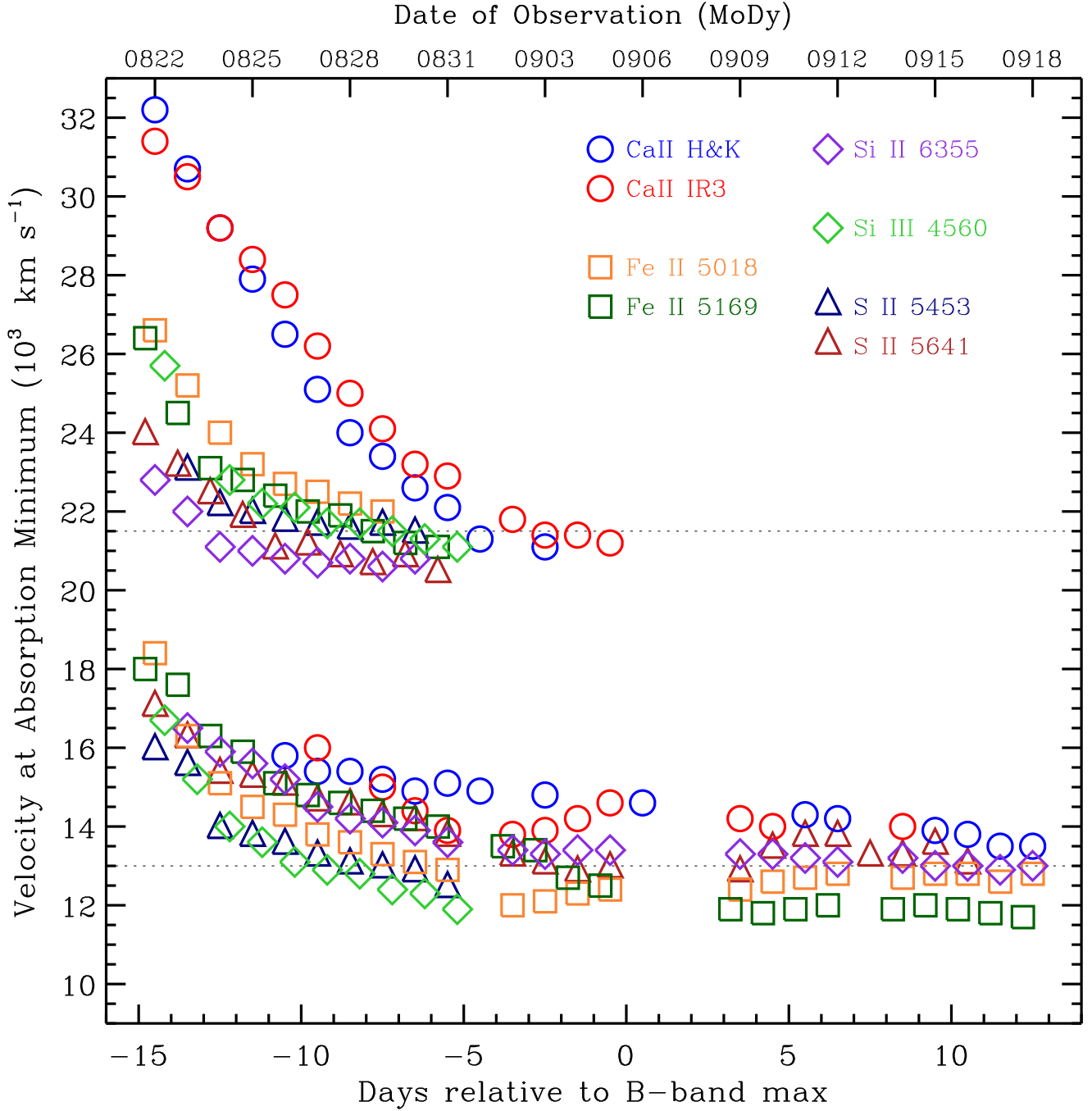}
\caption[]{Velocities of HVF and PVF are plotted by phase for 8 lines found in the spectra of SN~2009ig between $-14$~d and +13 d.  The measurements are listed in Tables~\ref{hvtbl} and \ref{pstbl}.  The lines and their identifying symbols are listed in the top-right corner.  Separate HVF and PVF line forming regions are evident with a gap of about 8,000 \kms\ at all phases.  No features are detected at velocities between the layers.  Measurement uncertainties are higher for HVF after $-7$~d and for PVF before $-12$~d (see Figure~\ref{gfit} and discussion in the text).  To reduce overplotting, data have been shifted slightly for \sithree\ \wl 4560 ($+0.3$~d), \s\ \wl 5641 ($-0.3$~d; HVF only), and \fe\ \wl 5169 ($-0.3$~d). \label{vplot}}
\end{figure*}

%%%%****** Figure 10 ******%%%%
\begin{figure}[t]
\center
%\hspace{-0.5cm} % To move left
\includegraphics[width=0.52\textwidth]{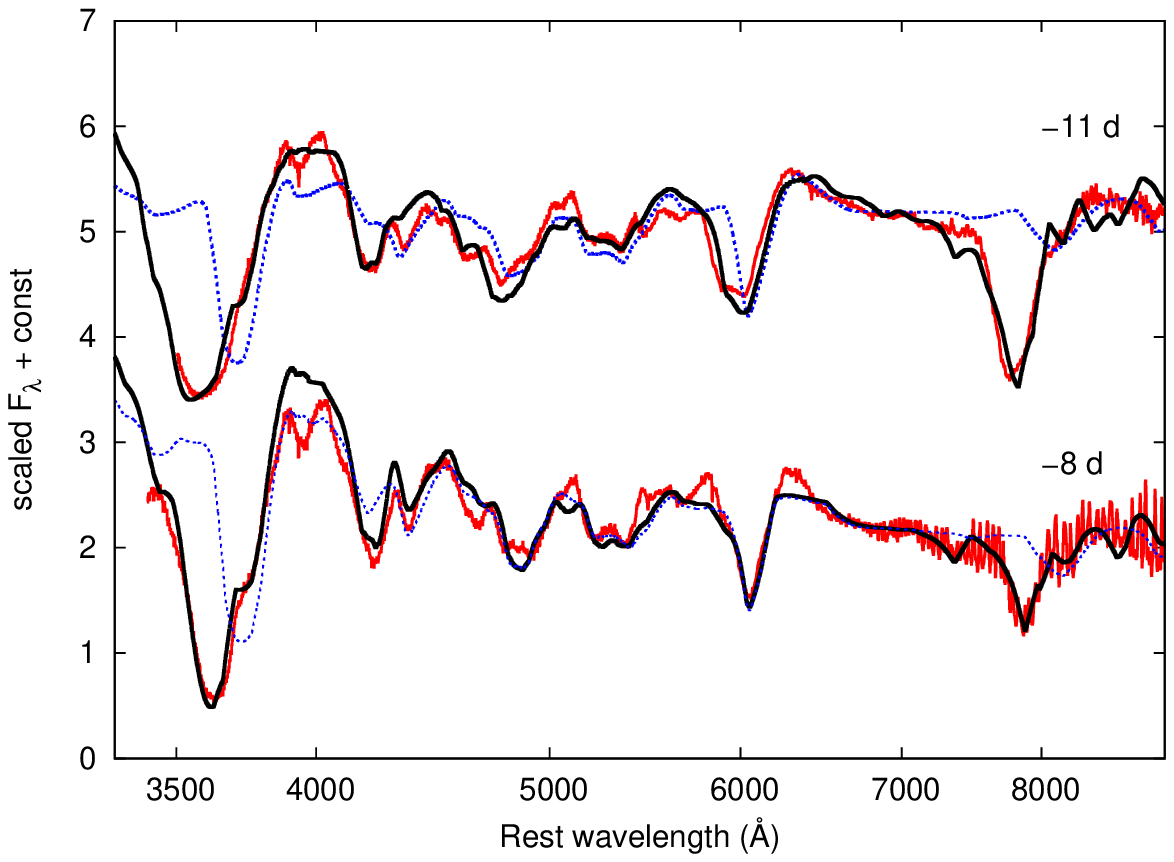}
\caption[]{Spectra from SN~2009ig obtained at $-11$~d and $-8$~d are plotted with spectra from SYNOW models.  The data are shown in red.  Model spectra with only PVF are dotted blue and model spectra with both PVF and HVF are solid black.  The addition of HVF lines clearly improves the agreement between the model and the data.}{ \label{syn}}
\end{figure}

%%%%%%%%%%%%%%%%%% Section %%%%%%%%%%%%%%%%%%%%%%
\section{The Location of HVF Line Forming Regions}
\label{lfr}

\placefigure{area4}
\placefigure{vplot}

Figure~\ref{area4} can be used to trace the location of the line forming regions for HVF of \ca\ H\&K, \ca\ IR3, \fe\ \wl 5169, and \si\ \wl 6355 at phases $-14$~d, $-12$~d, $-10$~d, $-6$~d, and $+0$~d (top to bottom).  Velocity serves as a proxy for radial distance, with the center of the SN on the right at 0 \kms\ and the radius increasing with velocity from right to left.  

The spectra are normalized to 1.0 at 6500 \AA\ and the features are normalized to a flat continuum.   The flux scale is the same in each panel.  At $-14$~d, all features have the same flux calibration.  At later phases, all other features retain the same flux calibration, but \ca\ H\&K HVF is scaled down to fit in the figure.

At $-14$~d, both \ca\ HVF (purple and red) occupy the same velocity space, and a blue wing is present in both features that extends from 40,000 to $\sim$55,000 \kms.  The minima for the \ca\ HVF are at velocities greater than 30,000 \kms\ (Table~\ref{hvtbl} and Figure~\ref{vplot}).  The HVF velocity from \fe\ \wl 5169 is near 26,000 \kms.  The line profile for \fe\ at this phase is similar to that of \ca\ H\&K in both width and depth.  A velocity difference of about 5,000 \kms\ is evident between the \ca\ line forming region (purple line) and \fe\ line forming region (blue line) on the red sides of the features.  The blue sides, however, are closely aligned.  This discrepancy is caused by the HVF of \fe\ \wl 5018 that creates a notch observed near 35,000 \kms\ in the velocity space of \fe\ \wl 5169 (blue line).  Between the red edge of this notch and the continuum, the side of the \fe\ \wl 5169 HVF is shifted by about 7,000 \kms\ to the blue.

The \si\ HVF at $-14$~d has a minimum at about 23,000 \kms, and the line forming region is confined to a velocity space that is almost entirely in the lower velocity half of the \fe\ HVF.  The \si\ line profile at $-14$~d (green) is clearly asymmetric, with a steeper blue side and an abrupt transition at the blue edge from the absorption to the continuum.  The blue limit of the \si\ HVF has nearly the same velocity as the absorption minimum for \ca.  In contrast, the PVF of \si\ ($-6$~d,$+0$~d) are nearly symmetric.

At $-12$~d, HVF are strong from all lines, with velocities for \ca\ about 29,000 \kms, for \fe\ near 23,000 \kms, and for \si\ near 21,000 \kms.  The \si\ HVF is again formed in a smaller region than for other lines.  The blue limit of the \si\ HVF is close to the same velocity as it was at $-14$~d.  PVF from \si\ and \fe\ can be detected near 16,000 \kms\ at this phase, with the \si\ PVF significantly distorting the line profile.  

At $-10$~d, the HVF of \si\ is noticeably weaker, but it is clearly present as it forms the distortion in the blue side of the \si\ PVF.  Using the methods described in \S \ref{id}, the velocity of HVF from \si\ is measured to be near 21,000 \kms.   The HVF velocity of \fe\ is near 22,000 \kms\ and the \ca\ HVF remains strong at about 27,000 \kms.  At this phase, the HVF of \fe\ \wl 5169 has a significant contribution from PVF \fe\ \wl 5018 that narrows the observed profile.  The \ca\ H\&K HVF also receives a strong contribution from PVF \si\ \wl 3858. 

At $-6$~d, the most obvious evidence for HVF is a very weak absorption from the \ca\ IR3 HVF.  The blue line absorption near 21,000 \kms\  is primarily due to the PVF of \fe\ \wl 5018.  The \ca\ H\&K HVF absorption (purple) near 23,000 \kms\ is a blend with PVF \si\ \wl 3858. The magnitude of the contribution from \si\ can be estimated by comparing the HVF of \ca\ H\&K to the HVF of \ca\ IR3.  The broad Ca/Si profile does not provide a HVF detection.  At this phase, \si\ has a PVF at 14,000 \kms, and the line profile is nearly symmetric in velocity space with no evidence of a HVF.  

Figure~\ref{area4} shows that the absorption minima for HVF are always observed above 20,000 \kms. There is no evidence at any phase for HVF with lower velocities.  By $+0$~d, all absorptions are easily explained by associations with PVF at velocities near 13,000 \kms.  The blue wing of the \ca\ IR3 PVF at this phase is slightly flattened by the remnant of the HVF.       
  
Figure~\ref{vplot} plots HVF and PVF velocity measurements by phase for \si\ \wl 6355, \sithree\ \wl 4560, \s\ \wl\wl 5453, 5641, \ca\ \wl\wl 3945, 8579, and \fe\ \wl\wl 5018, 5169.   All lines contribute both HVF and PVF measurements to the figure.  The plotted data are found in Tables~\ref{hvtbl} and \ref{pstbl}.  This figure is used to visually compare the relative velocities of both HVF and PVF from different lines at the same phases.  Variations in the decline rates are easy to discern.  HVF and PVF from the same lines or the same ions can also be compared.   

HVF velocities converge to a common value near 21,500 \kms\ and PVF velocities are grouped together at all phases.  A significant gap in velocity space is evident between the HVF and PVF regions for all lines at all phases.  The separation is about 8,000 \kms, except \ca\ which displays a greater separation before $-5$~d and \si\ \wl 6355 for which the separation at all phases is about 7,000 \kms.   No features are found at velocities that would place them between these layers. 

Figure~\ref{vplot} shows that from $-14$~d to $-10$~d, the HVF velocities of both \fe\ lines are intermediate between the higher \ca\ HVF velocities and the lower HVF velocities for other lines.  By $-9$~d, \fe\ HVF velocities are essentially the same as the velocities of other HVF that are not \ca.  

The patterns of velocity by phase can be compared the behavior of the HVF line profiles by phase that are displayed in Figure~\ref{area4}.   \si\ and \fe\ both have their maximum velocities in the earliest spectrum at $-14$~d.  At that phase, the \fe\ HVF line profile is wide, deep and more like the the HVF of \ca\ then the HVF of \si.  The velocity difference between the HVF of \fe\ and \si\ is 3,600 \kms.  At $-10$~d, the HVF line profiles are less distinct due to blending with PVF.  The \fe\ line profile is narrower and weaker and the velocities of \si\ and \fe\ are separated by 1,600 \kms.  The HVF of \ca\ at this phase remain broad and at higher velocities.  At $-6$~d, the HVF of \si\ and \fe\ are very weak and their measured velocities are within 400 \kms.

The velocity dispersion of the HVF components of \si, \sithree, \s, and \fe\ is confined within a narrow velocity region of less than 2000 \kms\ from $-12$~d until HVF are no longer detected.  By $-9$~d, the mean HVF velocity (excluding \ca) is about 21,500 \kms, and \ca\ velocities reach this value a few days later.  After the HVF velocities reach $\sim$ 21,500 \kms, the velocities remain constant within the range of scatter introduced by measurement uncertainties.  The mean PVF velocity is observed to be about 13,000 \kms\ at $-4$~d, and it declines less than 1,000 \kms\ through the end of our observations at $+13$~d.  (Most of the decline is caused by \fe\ \wl 5169; see \S \ref{ps}.)  The relative velocities of the HVF and PVF regions imply that, with homologous expansion, the radius of the HVF line forming region is $\sim 1.6$ times greater than the radius of the PVF line forming region.

Figure \ref{syn} shows SYNOW model spectra plotted with data from SN~2009ig obtained at $-11$~d and $-8$~d.  These phases are chosen because they include HVF and PVF, and the relative strengths of the components change in the time interval.  The data are plotted in the figure with red lines.  SYNOW spectra from models with only PVF velocities are plotted in blue with dotted lines, and model spectra with both HVF and PVF are solid black.  At both phases, the addition of HVF noticeably improves the models in the regions of the \ca\ and \si\ features.  More subtle improvements are evident around the \fe\ and \sithree\ features.

The SYNOW models using only PVF velocities produce a \ca\ IR3 feature that is weak and shifted to longer wavelengths because the \ca\ lines at $-11$~d and $-8$~d are dominated by the HVF component.  The relative weakness of the PVF \ca\ IR3 feature in the model is a consequence of SYNOW setting the optical depth for each ion to represent the strongest line (in this case \ca\ H\&K) and then calculating the strength of other lines from that ion assuming local thermodynamic equilibrium (LTE).   The SYNOW parameters for the models are given in Table \ref{syntbl}.  In these simple models, the same HVF and PVF velocities are assigned to all ions except for high-velocity \ca\ and \fe. 

Figure \ref{syn} can be compared to Figure~4 of \citet{Branch05}, which uses high-velocity \fe\ to improve the SYNOW fit to a $-12$~d spectrum from SN 1994D.  Additional HVF contributions from \mg\ and \ion{Fe}{3} can be added to further improve the agreement between SYNOW model spectra and the data, but HVF from these lines are not clearly identified in SN~2009ig.   \mg\ and \ion{Fe}{3} are not included in the velocity tables or plots.

%%%%****** Figure 11 ******%%%%
\begin{figure*}[t]
\center
\includegraphics[width=1.0\textwidth]{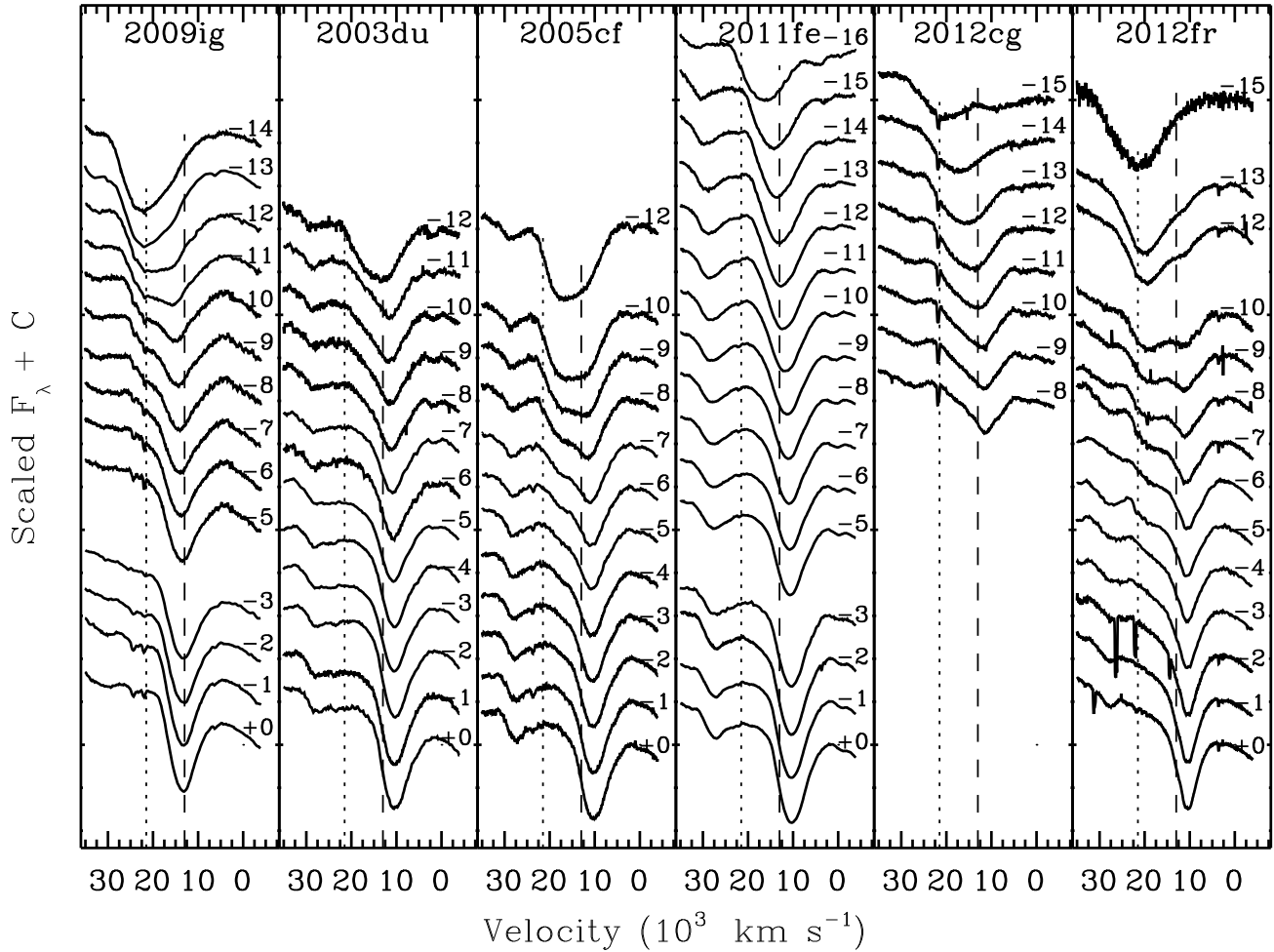}
\caption[]{The \si\ \wl 6355 absorption feature by phase in six SN~Ia, with the first observation at least 12 days before \bmax.  The earliest data clearly require separate HVF and PVF components to fit the line profiles, except for SN~2011fe which forms a single broad absorption feature.  The HVF and PVF components appear together for a few days.  After $-6$~d only the PVF is detected, and all spectra show similar line profiles with small differences between the SN in velocity space.  The sequence of line profiles from $-10$~d to $-5$~d in SN~2005cf is nearly identical to the $-12$~d to $-7$~d profiles from SN~2009ig.  All of the spectra in this figure can be correlated with a line profile in the sequence of SN~2009ig.  The dotted line at 21,500~\kms\ and the dashed line at 13,000~\kms\ are the same in each panel. \label{psi6}}
\end{figure*}

%%%%%%%%%%%%%%%%%% Section %%%%%%%%%%%%%%%%%%%%%%
\section{Comparison to HVF in Other SN I\lowercase{a}}
\label{comp}

\placefigure{psi6}

Figure~\ref{psi6} shows the development of \si\ \wl 6355 absorption features by phase in six SN~Ia for which there are high-cadence spectroscopic observations that begin at least 12 days before \bmax.\footnote{Some of the data in Figure~\ref{psi6} were obtained through the \emph{WISeREP} data archive \citep{Yaron12}.  The original published sources of the spectra are SN~2003du \citep{Stanishev07,Blondin12}, SN~2005cf \citep{Garavini07,Wang09a}, SN~2011fe \citep{Nugent11,Pereira13} (\emph{SNFactory}), and SN~2012fr  \citep{Childress13a}.  The SN~2012cg data are previously unpublished FAST spectra from the CfA archive.}  We note that none of these SN~Ia are rapid decliners; their $\Delta m_{15}(B)$ values (in SN order of discovery) are 1.02, 1.05, 0.90, 1.11, 0.86, and 0.80.

The spectra obtained before $-10$~d provide strong evidence for detached HVF regions from both \ca\ and \si\ in five of the six SN.  With one exception, the earliest spectra require separate HVF and PVF components to fit the line profiles.  Only SN~2011fe at $-16$~d has a broad absorption feature that can be fit with a single Gaussian.

The line profiles for every spectrum in Figure~\ref{psi6} can be correlated with a profile from one of the phases in the spectral sequence of SN~2009ig.  In some cases, the phase must be offset slightly to establish the best match.  For example, the sequence of line profiles from $-10$~d to $-5$~d in SN~2005cf is nearly identical to the $-12$~d to $-7$~d profiles in SN~2009ig.  The features of SN~2003du from $-12$~d to $-8$~d are similar to the features of SN~2009ig at the same phases, but the HVF diminish more rapidly in SN~2003du.  

The most extreme phase offset relative to SN~2009ig is found in SN~2011fe, for which the sequence from $-16$~d to $-10$~d corresponds to SN~2009ig from $-12$~d to $-6$~d.  Although the first spectrum of SN~2011fe was obtained earlier than the first spectrum of SN~2009ig with respect to \bmax, the spectrum of SN~2011fe corresponds to a phase in the sequence that is after the detached HVF are observed.  There is only a It is possible that even earlier observations of SN~2011fe could have revealed HVF.

Previous authors have studied the eccentricities of the \si\ \wl 6355 line profile in SN~Ia without directly detecting HVF.  The presence of a HVF component from \si\ \wl 6355 is inferred from unusual line profiles that are wide, square, or triangular.   None of the spectra available before SN~2009ig display separate and distinct high-velocity \si\ features such as found at $-14$ and $-13$~d in SN~2009ig and from $-14$~d to $-11$~d in SN~2012fr.  

\citet{Garavini07} use SYNOW to model early-time spectra of eight SN~Ia.  They show that fitting the \si\ \wl 6355 line profile requires a ``detached" \si\ region at 20,000--22,000 \kms\ for seven of the eight SN~Ia.  The other SN in their sample requires a detached region at 24,500 \kms.

\citet{Wang09a} compare the early spectra of six SN~Ia and report that each of the \si\ \wl 6355 line profiles are well fit by a double-Gaussian function with separate central wavelengths.  They show that two separate absorption features can be combined to produce flat-bottomed and triangular line profiles.  

Figure~\ref{psi6} shows that for different SN~Ia the characteristic velocities of the HVF and PVF regions can vary by up to 3,000 \kms, but in all cases a gap of at least 5,000 \kms\ is maintained between the HVF and PVF.  \citet{Garavini07}, \citet{Stanishev07}, \citet{Wang09a}, and \citet{Childress13a} describe the separation between HVF and PVF components to be about 8,000 \kms, the same value as the velocity gap in SN~2009ig.

The correlation of line profiles from pre-maximum spectra of other SN~Ia to the spectral sequence of SN~2009ig confirms a suggestion made by \citet{Stanishev07} that the ``peculiar'' profiles of \si\ \wl 6355 in early-time spectra of SN~Ia can be explained as part of a common evolutionary sequence that includes separate HVF and PVF components.  \citet{Stanishev07} also propose that the strength of the HVF components from \si\ and \ca\ lines is correlated in SN~Ia.  Thus, if one of these lines produces strong HVF, then the other will also have a strong HVF component.   The spectra of SN~2009ig confirm this relationship, and they also demonstrate that the presence of strong HVF from \si\ and \ca\ may be an indicator for HVF of \sithree, \s, and \fe. 

The agreement between the SN~2009ig results and these well-studied SN~Ia demonstrates that the behavior of HVF line forming regions in spectra of SN~2009ig is not unique, and in fact is quite common.  The papers discussed in this section identify fifteen SN~Ia with spectra obtained at $-10$~d or earlier.  All fifteen SN display high-velocity \si\ at velocities between 17,000 and 22,000 \kms, and all but SN~2011fe show evidence for separate HVF and PVF absorption components.

%%%%%%%%%%%%%%%%%% Section %%%%%%%%%%%%%%%%%%%%%%
\section{Possible Sources of HVF}
\label{sources}

The observations of detached HVF in SN~2009ig suggest the presence of a region at 20,000--23,000 \kms\ in the velocity space of the SN with a localized enhancement of element abundance or density, or both.  The proposed line forming region would be the source of HVF for \si, \s, and \fe\ from $-12$~d to $-5$~d.  

This simple model is complicated by the fact that HVF are initially detected at a range of even higher velocities. In addition, the HVF velocities of \si, \ca, and \fe\ decline at different rates.  This suggests that the lines may sample different density structures.  Although the velocity ranges overlap, the HVF of each line are not even detected in the same region.  Figure~\ref{area4} shows that at $-14$~d, \si\ \wl 6355 has the narrowest detached HVF with a velocity range from 10,000 to 30,000 \kms.  At the same phase, the HVF of \ca\ and \fe\ HVF extend from 10,000 to 50,000 \kms, but they are offset from each other.

\ca\ absorptions are exceptionally broad in velocity space, with absorption minima that are several thousand \kms\ higher than the HVF components of other lines.  Due to the very low excitation potentials of the \ca\ lines ($\le 1.7$ eV), they are capable of producing detectable absorptions at much lower abundances than most other lines found in SN~Ia.  \fe\ lines also have low excitation potentials ($\le 2.9$ eV).  In SN~2009ig, the HVF from \fe\ are nearly as wide and deep as HVF of \ca, with velocities between \ca\ and \si.  

The simplest explanation for both the HVF Ca and Fe observations is that primordial abundances of these elements are sufficient to produce the observed absorptions.  The line profiles can be produced by light from the SN passing through a line forming region that extends from about 10,000 to 50,000 \kms\ in velocity space.  The observed minima for these features would move to lower velocities with time, as expansion of the SN reduces opacity at high velocities more rapidly than it does at lower velocities.  As the column depth is reduced, the observed minima become nearly constant at velocities near to the inner edge of the line forming region.  

If the HVF of \ca\ and \fe\ are formed in regions of unburned material from the progenitor, it is still a challenge to explain the simultaneous presence of detached HVF for  \si\ \wl 6355 (8.1 eV) and \s\ (13.7 eV).  An explanation is also required for the fact that the HVF velocities for all lines, as measured at the absorption minima, eventually converge near 21,500 \kms.  Since Figure~\ref{area4} shows that the line forming materials are distributed through a wide range of velocities, there must be some motivation for the line opacities of \si, \sithree, \s, \ca, and \fe\ to have approximately equal values at the same phases and in a relatively narrow velocity range.  

Attempts have been made to explain HVF without invoking a separate line forming region.  The flat-bottomed and triangular shapes observed in the \si\ \wl 6355 line profile at early times receive particular attention.   None of the physical models used to generate unusual line profiles survive comparisons with early observations of multiple SN~Ia.  HVF from \ca\ have also been attributed to opacity effects from recombination, but the time scale of recombination is a day or two and HVF from \ca\ are observed from 8 to 15 days in individual SN~Ia.

Explosion models for SN~Ia make predictions for structures that range from spherical to highly asymmetric.  The resulting chemical distributions may be simply stratified by atomic mass, violently mixed between regions of burned material, or contain bubbles of material moving between burning regions.  Model results can be tuned to produce specific layering compositions and velocity distributions that may mimic HVF of specific atoms at specific locations.   We are not aware of any explosion models that predict the relative uniformity across a range of absorbing materials that is observed for HVF in SN~2009ig.  

One way to increase the local density in high-velocity regions is by interaction between the expanding ejecta and nearby circumstellar material \citep[CSM;][]{Gerardy04}.  The shock front sweeps up CSM and a shell is ``locked in" after the shock moves on.  The shell will continue in homologous expansion with the rest of the SN ejecta.  This model satisfies the constraint that the HVF line forming region exist in a narrow, unchanging velocity window and explains the asymmetric line profiles of the \si\ \wl\ 6355 HVF.

Three of the recent models have been able to produce variable density structures: symmetric ignitions that generate strong mixing \citep{Maeda10}, gravitationally confined detonations with burning regions on the surface of the WD \citep{Meakin09}, and double-detonation models that produce a high-velocity burning region with a density gap between it and the center of the SN \citep{Shen13}. 

%%%%%%%%%%%%%%%%%% Section %%%%%%%%%%%%%%%%%%%%%%
\section{Summary and Conclusions}
\label{conc}

We present a time series of spectra of SN~2009ig obtained between $-14.5$~d and $+12.5$~d with respect to the time of \bmax.   The two earliest observations in our sample ($-14.5$~d and $-13.5$~d) are the first spectra of a SN~Ia to resolve the high-velocity component of \si\ \wl 6355 as a distinctly separate absorption feature.   HVF are identified in the spectra from \si, \sithree, \s, \ca, and \fe, and we show that the line profiles require a detached HVF line forming region.

Simultaneous detections are made for HVF and PVF of the same lines from $-12$~d to $-5$~d.  Velocity measurements demonstrate that the HVF and PVF line forming regions are separated by $\sim 8,000$ \kms\ as long as HVF are detected.    Using a mean velocity of 21,500 \kms\ for the HVF layer and 13,000 \kms\ for the PVF layer, we show that $R_{\rm HVF} \approx 1.6 \, R_{\rm PVF}$.   

The initial HVF velocities and the velocity decline rates are different for each line.  The fact that the HVF velocities eventually converge suggests the presence of a line forming region in SN~20019ig at velocities between 20,000 and 23,000 \kms.  A comparison of the HVF from \si, \ca, and \fe\ by phase in velocity space shows that HVF from different lines are formed in overlapping but not identical regions.  HVF of \ca\ and \fe\ have strong features that range from 10,000 to 50,000 \kms\ in the earliest spectra, but the locations of their absorption minima differ by $\sim$ 7,000 \kms.  At the same phases, \si\ HVF are confined to a much narrower region that extends from 10,000 to only 30,000 \kms.  We explore the physical conditions that can produce detached HVF, and find that it is challenging for any model to reproduce the observations of SN~2009ig. 

The spectra of SN~2009ig form a complete map of the transition from the earliest observations of detached HVF to features with contributions from both HVF and PVF, and finally to phases when all of the absorptions are PVF.   HVF of both \ca\ and \si\ are frequent characteristics of SN~Ia observed before $-10$~d, and that allows us to compare our results to observations of several other SN~Ia.   All of the line profiles found in early-time spectra of SN~Ia can easily be correlated with individual profiles that are part of the development sequence defined by SN~2009ig.  We interpret this as evidence for a common evolutionary sequence in SN~Ia that requires both HVF and PVF line forming regions.

\acknowledgments

The authors wish to thank the Chairmen of the TACs from the University of Texas and Penn State University for providing Director's discretionary time for the HET/LRS observations. G.H.M. thanks Nick Suntzeff and Rob Robinson for insightful comments.  G.H.M. also thanks Mark Phillips for providing perspective, as well as Michael Childress for helpful discussions and for sharing an advance copy of his paper on SN~2012fr.  We thank Mark Sullivan, Isobel Hook, Peter Nugent, Andy Howell, and Bill Vacca for sharing their own data on line profiles in young SN~Ia.  The authors make frequent use of David Bishop's excellent webpage listing recent supernovae and valuable references associated with them: www.rochesterastronomy.org/snimages/.  J.V. is supported by Hungarian OTKA Grants K-76816 and NN-107637, NSF grant AST-0707769, and Texas Advanced Research Project ARP-009. The research of J.C.W. in supported in part by NSF grant AST-1109801.  The CfA Supernova Program is supported by NSF grant AST-1211196 to the Harvard College Observatory.  X. Wang is supported by the Natural Science Foundation of China (NSFC 11178003, 11073013), the China-973 Program 2009CB824800, and NSF grant AST-0708873 (through L. Wang).  E.Y.H. is supported by NSF grant AST-1008343. A.V.F. is grateful for financial assistance from NSF grant AST-1211916, the TABASGO Foundation, and the Christopher R. Redlich Fund. We greatly appreciate the valuable assistance provided by staff members at the observatories where SN~2009ig was observed.

{\it Facilities:}  \facility{FLWO:1.5m (FAST)}, \facility{HET (LRS)}, \facility{Keck:I (LRIS)}, \facility{MMT (Blue Channel)}, \facility{Lick:Shane (Kast)}, \facility{TNT}, \facility{Swift (UVOT; UV grism)}

%%%%%%%%%%%%%%%%%% Bibliography %%%%%%%%%%%%%%%%%%%%%%

%%%%%%%%%%%%%%%%%%%%%%%%%%%
% Tables
%%%%%%%%%%%%%%%%%%%%%%%%%%%

\clearpage
\center

\begin{deluxetable}{rrccc}
\tabletypesize{\scriptsize}
\tablecolumns{5} 
\tablewidth{0pc}
\tablecaption{Spectroscopic Observations\label{obstbl}}
\tablehead{\colhead{Date}  & \colhead{Phase\tablenotemark{a}} & \colhead{Telescope/Instrument} & \colhead{Range\tablenotemark{b} (\AA)} & \colhead{$\lambda$ Merge\tablenotemark{c} (\AA)}}

\startdata
Aug. 22.5   &  -14.5   &   Lick/Kast  & 3400--9000 & 6700 \\
Aug. 22.6   &  -14.4   &   Keck I/LRIS  & 3100--7860 & 6700 \\
Aug. 23.4   &  -13.6  &   HET/LRS         & 4210--9000 & 4225 \\
Aug.  23.6   &  -13.4   &   Swift/Ugrism  & 3200--4100 & 4070 \\
Aug. 24.5   &  -12.5   &   Lick/Kast  & 3500--9000 & \nodata \\
Aug. 25.5   &  -11.5   &   Lick/Kast  & 3500--9000 & 4000 \\
Aug.  25.6   &  -11.4   &   Swift/Ugrism  & 3200--4400 & 4000 \\
Aug. 26.5   &  -10.5   &   MMT/Blue Channel  & 3190--8330 & \nodata  \\
Aug. 27.5   &  -9.5   &   Lick/Kast  & 3400--9000 & 6700 \\
Aug. 27.5   &  -9.5   &   MMT/Blue Channel  & 3180--8270 & 6700 \\
Aug. 28.5   &  -8.5   &   Lick/Kast  & 3400--9000  & 6700 \\
Aug. 28.5   &  -8.5   &   MMT/Blue Channel  & 3180--8320  & 6700 \\
Aug. 29.4   &  -7.6   &   HET/LRS         & 7170--9000  & 7210 \\
Aug. 29.5   &  -7.5   &   MMT/Blue Channel  & 3190--8270  & 7210 \\
Aug. 30.4   &  -6.6   &   HET/LRS         & 4710--9000  & 6850 \\
Aug. 30.4   &  -6.6   &   MMT/Blue channel  & 3180--8250  & 6850 \\
Aug. 31.4   &  -5.6   &   HET/LRS         & 4170--9000  & \nodata \\
Sep. 1.6    &  -4.4   &   Swift/Ugrism  & 3200--4900  & \nodata \\
Sep. 2.4    &  -3.6   &   HET/LRS         & 4160--9000  & \nodata \\
Sep. 3.6    &  -2.4   &   Swift/Ugrism  & 3200--4400  & 4200 \\
Sep. 3.5    &  -2.5   &   HET/LRS         & 4190--9000  & 4200 \\
Sep. 4.4    &  -1.6   &   HET/LRS         & 4180--9000  & \nodata \\
Sep. 5.4    &  -0.6   &   HET/LRS         & 4180--9000  & \nodata \\
Sep. 6.7   &  +0.7   &   Swift/Ugrism  & 3200--4900  & \nodata \\
Sep. 9.4    &  +3.4   &   HET/LRS         & 4180--9000  & \nodata \\
Sep. 10.4  &  +4.4   &   HET/LRS         & 4180--9000  & \nodata \\
Sep. 11.5  &  +5.5   &   FLWO/FAST   & 3300--7200  & \nodata \\
Sep. 12.5  &  +6.5   &   FLWO/FAST   & 3300--7200  & \nodata \\
Sep. 14.4  &  +8.4   &   HET/LRS         & 4180--9000  & \nodata \\
Sep. 15.5  &  +9.5   &   FLWO/FAST   & 3300--7200  & \nodata \\
Sep. 16.5  &  +10.5   &   FLWO/FAST   & 3300--7200  & \nodata \\
Sep. 17.4  &  +11.5   &   FLWO/FAST   & 3300--7200  & \nodata \\
Sep. 18.5  &  +12.5   &   FLWO/FAST   & 3300--7200  & \nodata \\
\enddata
\tablecomments{The spectra obtained before Sep. 9 were previously published by \citet{Foley12}.}
\tablenotetext{a}{Phase in days with respect to the time of $B$-max (UT 2009 Sep 6.0 = JD 2,455,080.5).}
\tablenotetext{b}{Wavelength range used for this paper.  The complete spectra may cover a larger range.}
\tablenotetext{c}{Wavelength at which this spectrum was trimmed and combined with a contemporaneous spectrum to form a single spectrum for this date.}

\end{deluxetable}

%%%%%%%%%%%%%%%%%%%%%%%%%%%%%%%%%%%%%%%%%%%%%%%%%%%%%%
% HVF measurements

\begin{deluxetable}{crcccccccc}
\tabletypesize{\scriptsize}
\tablecolumns{10} 
\tablewidth{0pc}
\tablecaption{High-Velocity Features in SN~2009\lowercase{ig} ($10^3$ \kms)\tablenotemark{a}\label{hvtbl}}
\tablehead{\colhead{Date} & \colhead{Phase\tablenotemark{b}} & \colhead{Ca~II} & \colhead{Si~III} & \colhead{Fe~II} &
 \colhead{Fe~II} &   \colhead{S~II} &  \colhead{S~II} & \colhead{Si~II} & \colhead{Ca~II} \\
 \colhead{} & \colhead{} & \colhead{$\lambda$3945} & \colhead{$\lambda$4560} & \colhead{$\lambda$5018} & 
 \colhead{$\lambda$5169} &  \colhead{$\lambda$5453} &  \colhead{$\lambda$5641} & \colhead{$\lambda$6355} & \colhead{$\lambda$8579}}
\startdata
Aug. 22 &  -14 &  32.2 &  25.7 &  26.6 &  26.4 &   \nodata &  23.9 &  22.8 &  31.4 \\ 
Aug. 23 &  -13 &  30.7 &   \nodata &  25.2 &  24.5 &  23.0 &  23.1 &  22.0 &  30.5 \\ 
Aug. 24 &  -12 &  29.2 &  22.8 &  24.0 &  23.1 &  22.1 &  22.4 &  21.1 &  29.2 \\ 
Aug. 25 &  -11 &  27.9 &  22.2 &  23.2 &  22.8 &  21.9 &  21.8 &  21.0 &  28.4 \\ 
Aug. 26 &  -10 &  26.5 &  22.1 &  22.7 &  22.4 &  21.7 &  21.0 &  20.8 &  27.5 \\ 
Aug. 27 &   -9 &  25.1 &  21.7 &  22.5 &  22.0 &  21.6 &  21.1 &  20.7 &  26.2 \\ 
Aug. 28 &   -8 &  24.0 &  21.7 &  22.2 &  21.9 &  21.5 &  20.8 &  20.8 &  25.0 \\ 
Aug. 29 &   -7 &  23.4 &  21.5 &  22.0 &  21.5 &  21.6 &  20.6 &  20.6 &  24.1 \\ 
Aug. 30 &   -6 &  22.6 &  21.3 &   \nodata &  21.2 &  21.4 &  20.8 &  20.8 &  23.2 \\ 
Aug. 31 &   -5 &  22.1 &  21.1 &   \nodata &  21.1 &   \nodata &  20.4 &   \nodata &  22.9 \\ 
Sep. 01 &   -4 &  21.3 &  21.0 &   \nodata &   \nodata &   \nodata &   \nodata &   \nodata &   \nodata \\ 
Sep. 02 &   -3 &   \nodata &   \nodata &   \nodata &   \nodata &   \nodata &   \nodata &   \nodata &  21.8 \\ 
Sep. 03 &   -2 &  21.1 &   \nodata &   \nodata &   \nodata &   \nodata &   \nodata &   \nodata &  21.4 \\ 
Sep. 04 &   -1 &   \nodata &   \nodata &   \nodata &   \nodata &   \nodata &   \nodata &   \nodata &  21.4 \\ 
Sep. 05 &    0 &   \nodata &   \nodata &   \nodata &   \nodata &   \nodata &   \nodata &   \nodata &  21.2 \\ 
\enddata
\tablecomments{Measurement uncertainties are higher for HVF after $-7$~d (see discussion in text).} 
\tablenotetext{a}{Throughout this paper we represent velocities for blueshifted absorption features with positive values.}
\tablenotetext{b}{Rise time $\tau_r = 17.1$~d \citep{Foley12}, so the first spectrum was obtained 2.6~d after the explosion.}
\end{deluxetable}

%%%%%%%%%%%%%%%%%%%%%%%%%%%%%%%%%%%%%%%%%%%%%%%%%%%%%%
% PVF measurements

\begin{deluxetable}{crcccccccc}
\tabletypesize{\scriptsize}
\tablecolumns{10} 
\tablewidth{0pc}
\tablecaption{Photospheric Features in SN~2009\lowercase{ig}  ($10^3$ \kms)\tablenotemark{a}\label{pstbl}}
\tablehead{\colhead{Date} & \colhead{Phase\tablenotemark{b}} & \colhead{Ca~II} & \colhead{Si~III} & \colhead{Fe~II} &
 \colhead{Fe~II} &   \colhead{S~II} &  \colhead{S~II} & \colhead{Si~II} & \colhead{Ca~II} \\
 \colhead{} & \colhead{} & \colhead{$\lambda$3945} & \colhead{$\lambda$4560} & \colhead{$\lambda$5018} & 
 \colhead{$\lambda$5169} &  \colhead{$\lambda$5453} &  \colhead{$\lambda$5641} & \colhead{$\lambda$6355} & \colhead{$\lambda$8579}}
\startdata
Aug. 22 &  -14 &   \nodata &  16.7 &  18.4 &  18.0 &  15.9 &  17.0 &   \nodata &   \nodata \\ 
Aug. 23 &  -13 &   \nodata &  15.2 &  16.3 &  17.6 &  15.5 &  16.2 &  16.5 &   \nodata \\ 
Aug. 24 &  -12 &   \nodata &  14.0 &  15.1 &  16.3 &  13.9 &  15.3 &  15.9 &   \nodata \\ 
Aug. 25 &  -11 &   \nodata &  13.6 &  14.5 &  15.9 &  13.7 &  15.2 &  15.6 &   \nodata \\ 
Aug. 26 &  -10 &  15.8 &  13.1 &  14.3 &  15.1 &  13.5 &  15.0 &  15.2 &   \nodata \\ 
Aug. 27 &   -9 &  15.4 &  12.9 &  13.8 &  14.8 &  13.2 &  14.6 &  14.5 &  16.0 \\ 
Aug. 28 &   -8 &  15.4 &  12.8 &  13.6 &  14.6 &  13.0 &  14.5 &  14.2 &   \nodata \\ 
Aug. 29 &   -7 &  15.2 &  12.4 &  13.3 &  14.4 &  12.9 &  14.2 &  14.1 &  15.0 \\ 
Aug. 30 &   -6 &  14.9 &  12.3 &  13.1 &  14.2 &  12.8 &  14.1 &  13.9 &  14.4 \\ 
Aug. 31 &   -5 &  15.1 &  11.9 &  12.9 &  14.0 &  12.4 &  13.7 &  13.6 &  13.9 \\ 
Sep. 01 &   -4 &  14.9 &  12.1 &   \nodata &   \nodata &   \nodata &   \nodata &   \nodata &   \nodata \\ 
Sep. 02 &   -3 &   \nodata &   \nodata &  12.0 &  13.5 &   \nodata &  13.2 &  13.4 &  13.8 \\ 
Sep. 03 &   -2 &  14.8 &   \nodata &  12.1 &  13.4 &   \nodata &  13.0 &  13.3 &  13.9 \\ 
Sep. 04 &   -1 &   \nodata &   \nodata &  12.3 &  12.7 &   \nodata &  12.8 &  13.4 &  14.2 \\ 
Sep. 05 &    0 &   \nodata &   \nodata &  12.4 &  12.5 &   \nodata &  12.9 &  13.4 &  14.6 \\ 
Sep. 06 &    1 &  14.6 &   \nodata &   \nodata &   \nodata &   \nodata &   \nodata &   \nodata &   \nodata \\ 
Sep. 07 &    2 &   \nodata &   \nodata &   \nodata &   \nodata &   \nodata &   \nodata &   \nodata &   \nodata \\ 
Sep. 08 &    3 &   \nodata &   \nodata &   \nodata &   \nodata &   \nodata &   \nodata &   \nodata &   \nodata \\ 
Sep. 09 &    4 &   \nodata &   \nodata &  12.4 &  11.9 &   \nodata &  12.8 &  13.3 &  14.2 \\ 
Sep. 10 &    5 &   \nodata &   \nodata &  12.6 &  11.8 &   \nodata &  13.4 &  13.3 &  14.0 \\ 
Sep. 11 &    6 &  14.3 &   \nodata &  12.7 &  11.9 &   \nodata &  13.7 &  13.2 &   \nodata \\ 
Sep. 12 &    7 &  14.2 &   \nodata &  12.8 &  12.0 &   \nodata &  13.7 &  13.1 &   \nodata \\ 
Sep. 13 &    8 &   \nodata &   \nodata &   \nodata &   \nodata &   \nodata &  13.2 &   \nodata &   \nodata \\ 
Sep. 14 &    9 &   \nodata &   \nodata &  12.7 &  11.9 &   \nodata &  13.2 &  13.2 &  14.0 \\ 
Sep. 15 &   10 &  13.9 &   \nodata &  12.8 &  12.0 &   \nodata &  13.5 &  13.0 &   \nodata \\ 
Sep. 16 &   11 &  13.8 &   \nodata &  12.8 &  11.9 &   \nodata &  13.0 &  13.0 &   \nodata \\ 
Sep. 17 &   12 &  13.5 &   \nodata &  12.6 &  11.8 &   \nodata &   \nodata &  12.9 &   \nodata \\ 
Sep. 18 &   13 &  13.5 &   \nodata &  12.8 &  11.7 &   \nodata &   \nodata &  13.0 &   \nodata \\ 
\enddata
\tablecomments{Measurement uncertainties are higher for PVF before $-12$~d (see discussion in text).} 
\tablenotetext{a}{Throughout this paper we represent velocities for blueshifted absorption features with positive values.}
\tablenotetext{b}{Rise time $\tau_r = 17.1$~d \citep{Foley12}, so the first spectrum was obtained 2.6~d after the explosion.}
\end{deluxetable}

%%%%%%%%%%%%%%%%%%%%%%%%%%
% SYNOW Table
\begin{deluxetable}{lcccccccc}
\tabletypesize{\scriptsize}
\tablecolumns{9} 
\tablewidth{0pc}
\tablecaption{SYNOW Parameters for Model Spectra \label{syntbl}}
\tablehead{ \colhead{Ion} &  \colhead{$\tau$} &  \colhead{$v_{\rm min}$} & 
 \colhead{$v_{\rm max}$} &  \colhead{$v_e$} &  \colhead{$\tau$} &  
 \colhead{$v_{\rm min}$} &  \colhead{$v_{\rm max}$} &  \colhead{$v_e$} \\
\multicolumn{5}{c}{$-11$~day} & \multicolumn{4}{c}{$-8$~day} \\
\multicolumn{5}{c}{$v_{\rm phot}$ = 16,000 km s$^{-1}$} & \multicolumn{4}{c}{$v_{\rm phot}$ = 13,800 km s$^{-1}$}}
\startdata
Si II  & 3.0 & 16 & 30 & 2.0 & 1.2 & 16 & 30 & 2.0 \\ 
Si II(HVF) & 1.0 & 23 & 50 & 2.0 & 0.1 & 23 & 50 & 2.0 \\  
Si III & 0.8 & 16 & 50 & 2.0 & 0.8 & 13 & 50 & 2.0 \\
Si III(HVF) & 0.5 & 23 & 50 & 2.0 & 0.5 & 23 & 50 & 2.0 \\
Mg II(HVF) & 0.5 & 22 & 50 & 4.0 & 0.3 & 22 & 50 & 4.0 \\
S II & 0.7 & 16 & 50 & 2.0 & 0.5 & 13 & 50 & 2.0 \\
S II(HVF) & 0.1 & 23 & 50 & 2.0 & 0.1 & 23 & 50 & 2.0 \\
Ca II & 2.0 & 16 & 50 & 8.0 & 2.0 & 13 & 50 & 8.0 \\
Ca II(HVF) & 25 & 28 & 50 & 4.0 & 6.0 & 26 & 50 & 4.0 \\
Fe II(HVF) & 0.5 & 28 & 50 & 2.0 & 0.05 & 23 & 50 & 2.0 \\
Fe III & 0.5 & 16 & 50 & 2.0 & 0.5 & 16 & 50 & 2.0 \\

\enddata

\tablecomments{$\tau$: reference line optical depth; $v_{\rm min}$: minimum velocity;
$v_{\rm max}$: maximum velocity; $v_e$: $e$-folding velocity of optical depth profile. All
velocities are in units of $10^3$ km s$^{-1}$. Both models assume $T_{\rm eff} = 12,000$ K.}

\end{deluxetable}

%%%%%%%%%%%%%%%%%%%%%%%%%%

\end{document}